\pdfoutput=1
\documentclass[a4paper,12pt]{article}
\usepackage{graphicx,latexsym,color,dsfont}
\usepackage{amsmath,amsfonts,amsbsy,amssymb,amsthm}
\usepackage[utf8]{inputenc}
\usepackage[mathscr]{euscript}
\usepackage[T1]{fontenc}
\usepackage{mathtools,physics}
\usepackage[english]{babel}
\usepackage{scalerel}
\usepackage[medium]{titlesec}
\usepackage{bm}
\usepackage{cite}
\usepackage[hmargin=.7in,vmargin=1.1in]{geometry}
\usepackage[bookmarksnumbered=true,bookmarksopen=true]{hyperref}
\hypersetup{colorlinks,
	linkcolor=[rgb]{0,0.2,0.6},
	citecolor=[rgb]{0,0.2,0.6}}

\newcommand{\be}{\begin{eqnarray}}
\newcommand{\ee}{\end{eqnarray}}

\begin{document}
	
\title{Gauge Invariance, Polar Coordinates and Inflation}
\author{Ali Akil$^{a,b,c}$\footnote{Email: aakil@connect.ust.hk}{},~~ Xi Tong$^{a,b}$\footnote{Email: xtongac@connect.ust.hk}{}\\[2mm]
	\normalsize{$^a$\emph{Department of Physics, The Hong Kong University of Science and Technology,}}\\
	\normalsize{\emph{Clear Water Bay, Kowloon, Hong Kong, P.R.China}}\\
	\normalsize{$^b$\emph{Jockey Club Institute for Advanced Study, The Hong Kong University of Science and Technology,}}\\
	\normalsize{\emph{Clear Water Bay, Kowloon, Hong Kong, P.R.China}}\\
	\normalsize{$^c$\emph{Department of Physics, Southern University of Science and Technology (SUSTech),}}\\
	\normalsize{\emph{ Shenzhen 518055, P.R.China}}
}

\date{}	
\maketitle

\begin{abstract}
	We point out the necessity of resolving the apparent gauge dependence in the quantum corrections of cosmological observables for Higgs-like inflation models. We highlight the fact that this gauge dependence is due to the use of an asymmetric background current which is specific to a choice of coordinate system in the scalar manifold. Favoring simplicity over complexity, we further propose a practical shortcut to gauge-independent inflationary observables by using effective potential obtained from a polar-like background current choice. We demonstrate this shortcut for several explicit examples and present a gauge-independent prediction of inflationary observables in the Abelian Higgs model. Furthermore, with Nielsen's gauge dependence identities, we show that for any theory to all orders, a gauge-invariant current term gives a gauge-independent effective potential and thus gauge-invariant inflationary observables.
\end{abstract}

\maketitle
\section{Introduction}
The effective potential was introduced as an attempt to include the quantum corrections to the tree level potential, after which one can treat the background evolution in quantum field theory using classical variational methods \cite{Coleman:1973jx,Dolan:1973qd}. It is widely applied to the study of phenomena of spontaneous symmetry breaking. For example, the effective potential plays an important role in the determination of the quantum vacuum \cite{Coleman:1973jx} and its decay rate \cite{Linde:1981zj,Metaxas:1995ab}, the analysis of cosmological phase transitions \cite{Weinberg:1974hy,Linde:1978px,Quiros:1999jp,Patel:2011th,Garny:2012cg}, as well as inflationary dynamics \cite{Linde:1981mu,Guth:1982ec,Brandenberger:1984cz,Cook:2014dga}.
	
However, as an off-shell quantity by itself, the effective potential was shown to suffer from the problem of gauge dependence \cite{Jackiw:1974cv}.  First, it was argued in \cite{Dolan:1973qd} that the unitary gauge is the right way to extract physical quantities, being only a field redefinition and arguably not really a choice of gauge \cite{Higgs:1966ev,Kibble:1967sv}. Then, by means of Ward-Takahashi identity of BRST symmetry, \cite{Nielsen:1975fs} derived a first order partial differential equation (later improved in \cite{Aitchison:1983ns}) describing the exact dependence of the effective potential $V_{\text{eff}}(\bar{\phi},\xi)$ on the gauge parameter $\xi$,
\begin{equation}\label{NielsenId}
	\left(\xi\frac{\partial}{\partial\xi} -C(\bar{\phi},\xi)\frac{\partial}{\partial\bar{\phi}}\right)V_{\text{eff}}(\bar{\phi},\xi)=0~.
\end{equation}
It is argued that the physical states of the Hilbert space are represented by the different characteristic lines of the partial differential equation (\ref{NielsenId}). Gauge transformations map each characteristic line to itself and therefore leaves the physical states invariant.
	
It was later found in \cite{Tye:1996au} that if one expresses the scalar field in the polar field space coordinates, the gauge dependence cancels out and, interestingly, the result is in exact accordance with the effective potential in the unitary gauge with Cartesian field coordinates, thus matching the conclusions of \cite{Dolan:1974gu}. For a $U(1)$-symmetric gauge theory, it is to be expected that polar-coordinate fields give a clearer picture of physics since the symmetry of the theory is made manifest. Naturally, after choosing field space coordinates as fundamental fields, corresponding source terms proportional to these fundamental fields must be introduced into the partition function, so as to lead the system off-shell. 
It then becomes clear that if one excites the system using a symmetric source term, the off-shell effective action is automatically gauge-invariant even non-perturbatively, whereas an asymmetric source term will inevitably cause the off-shell effective action to pick up gauge-dependence.
	
One might question the necessity of requiring gauge-invariance for off-shell quantities like the effective potential, since processes are physical only when the background current source is removed and the system is put on-shell. For purposes of calculating particle scattering amplitudes around the vacuum, removing the source is equivalent to sending the field background to the quantum vacuum, where $\frac{\partial V_{\text{eff}}}{\partial\bar{\phi}}=0$. Then the Nielsen identity (\ref{NielsenId}) shows that at this minimum of the effective potential, the potential value is independent of gauge choices. Similar identities show that the on-shell scattering amplitudes are also independent of $\xi$ at the minimum. However, this is not the whole story, since in general, the field background can have a non-trivial spacetime dependence. For this purpose of an out-of-equilibrium process, removing the background current is not simply setting the system to vacuum but letting the background to evolve freely according to the variational principle of the effective action. The gauge dependence issue in this scenario is much more subtle. 
	
Take the example of inflation, if the inflaton field is not a gauge singlet, its classical background rolls according to the effective potential which receives gauge-dependent contributions from quantum fluctuations. As a consequence, inflationary observables such as slow-roll parameters and e-folding numbers appear to be polluted by a gauge-dependent quantum corrections. Since physical observables should not depend on gauge choices, this apparent gauge dependence must be resolved. 
We will argue that either one has to go to the non-perturbative regime \cite{Pimentel:2019otp} or has to analyze the slow-roll dynamics more carefully using in-in formalism. In the literature there are preliminary attempts \cite{George:2012xs,George:2013iia} to advance in this direction, but to our knowledge, no explicit demonstration of gauge-invariant inflationary observables has yet been made in this formalism. 

Among the other attempts, \cite{Jacoppo1,Jacoppo2} rely on the absorption of the anomalous dimension
of the Higgs field in the context of the Higgs inflation. Similar techniques also appear in \cite{Espinosa, Urbano}. However, we point out that their calculations are performed in the non-linear $R_\xi$ gauge family, and that the gauge dependence cancellation in their methods fails in the linear $R_\xi$ gauges. This mismatch between different gauge families is itself a form of unphysical gauge dependence, one that cannot be resolved by field canonicalization\footnote{See Appendix.~\ref{AHiggsInfAppendix} for a more detailed discussion.}. Meanwhile, \cite{Espinosa,Nielsen:2014spa} try to follow the characteristic lines of Nielsen equation and perform a corresponding field redefinition. This method, seems to us, still suffers an initial-gauge dependence. Because the solution is trying to adjust itself accordingly after a change of $\xi$, such that the observables stay invariant. However, we could have chosen a different $\xi$ initially and applied the same method to let it stay invariant. The observables will still disagree between different initial choices.

In favor of \textit{simplicity} over a heavy machinery of in-in calculations or even non-perturbative treatments, we propose the second possibility of resolution. By using a symmetric background current which fits naturally with polar field coordinates \cite{Tye:1996au}, the system is lifted up in a gauge-invariant way. And the resulting effective action is manifestly independent of gauge choices even off-shell. Then by setting the background current to zero, one obtain a gauge-independent description of background evolution, along with gauge-independent inflationary observables.
	
This paper is organized as follows. In Sect.~\ref{Cosmo}, we show the impact of apparent gauge dependence on inflationary observables and argue the necessity of resolving this issue. In Sect.~\ref{PolarExamps}, we identify the cause for this apparent gauge dependence and propose a simple and practical solution by using polar-like coordinates. The examples of $SU(2)$ in both the fundamental and the adjoint representations as well as $SU(2) \times U(1)$ are given to illustrate this idea. For the $U(1)$ model, we also present the predictions of inflationary observables using our method. Then in Sect.~\ref{GaugeDepId}, we use Nielsen's gauge dependence identity to give a non-perturbative proof of the off-shell gauge invariance of the polar-coordinate effective action. At last, in Sect.~\ref{unitaryLim}, we discuss the relation between polar coordinates and the unitary gauge. We point out the limitations of our method and conclude in Sect.~\ref{conclusion}.
	
\section{Gauge dependence problem in inflation}\label{Cosmo}
Inflation is the leading paradigm of early universe. In simplest setups, it is the exponential expansion period of the universe driven by a single slowly rolling scalar field called the inflaton. In principle, other quantum fields are also present during inflation. These fields usually have vanishing classical backgrounds and do not directly influence the background evolution. However, if they interact with the inflaton, quantum fluctuations of these fields will contribute to the effective potential of the inflaton. These are the radiative corrections to the background evolution. In particular, consider an Abelian Higgs inflation model \cite{Bezrukov:2007ep},
\begin{equation}
	S[g_{\mu\nu},\phi,A_\mu]=\int\sqrt{-g}\left[-\frac{M_p^2 R}{2}-\alpha |\phi|^2R+|D\phi|^2-V(|\phi|)-\frac{1}{4}F^2\right]~,
\end{equation}
where $D=\partial+ie A$ is the covariant derivative and $V(|\phi|)=-m^2|\phi|^2+\lambda|\phi|^4$ is the tree-level Higgs potential. The second term is a non-minimal coupling between gravity and the radial gauge singlet component of the Higgs field. This term naturally arises as a back-reaction of fluctuations of matter field $\phi$ to the spacetime geometry. The gauge field fluctuations contribute to both the gravitational sector and to the scalar sector. Since the gauge field does not acquire a classical background, we neglect its back-reaction to the gravity sector and only consider the back-reaction to the scalar sector, which can be accounted for by replacing the classical action by the gauge-fixed effective action computed in the spacetime background of $g_{\mu\nu}$,
\begin{equation}
	S_{\text{eff}}[g_{\mu\nu},\bar{\phi}_J,\xi]=\int\sqrt{-g}\left[-\frac{M_p^2 R}{2}-\alpha |\bar{\phi}_J|^2R\right]+\Gamma[\bar{\phi}_J,\xi]|_{g_{\mu\nu}}~.
\end{equation}
Since the quantum corrections come from fluctuations deep in the UV, we expect the spacetime curvature to play only a subdominant role for the fluctuating matter fields and approximate 
\begin{equation}\label{gToEta}
	\Gamma[\bar{\phi}_J,\xi]|_{g_{\mu\nu}}\approx\Gamma[\bar{\phi}_J,\xi]|_{\eta_{\mu\nu}}~.
\end{equation}
Therefore, when truncated to the second order in field gradients, the action looks like
\begin{equation}\label{JorEffAction}
	S_{\text{eff}}[g_{\mu\nu},\bar{\phi}_J,\xi]=\int\sqrt{-g}\left[-\frac{M_p^2 R}{2}-\alpha_{\text{eff}}(\bar{\phi}_J,\xi) |\bar{\phi}_J|^2R+Z(\bar{\phi}_J,\xi)|\partial\bar{\phi}_J|^2-V_{\text{eff}}(\bar{\phi}_J,\xi)\right]~.
\end{equation}
The above expressions are all written in the Jordan frame. To obtain an inflationary solution, one usually goes to the Einstein frame by performing a conformal transformation
\begin{equation}
	\hat{g}_{\mu\nu}=\Omega^2 g_{\mu\nu},~~~\chi_J=\int \sqrt{2Z\frac{\Omega^2+12\alpha_{\text{eff}}^2\bar{\phi}_J^2/M_p^2}{\Omega^4}}d\bar{\phi}_J,~~\text{where }~~~\Omega^2=1+\frac{2\alpha_{\text{eff}} \bar{\phi}_J^2}{M_p^2}~.\label{ZJordanToEinstein}
\end{equation}
Finally, the action in Einstein frame becomes
\begin{equation}\label{EinEffAction}
	S_{\text{eff}}[\hat{g}_{\mu\nu},\chi_J,\xi]=\int\sqrt{-\hat{g}}\left[-\frac{M_p^2 \hat{R}}{2}+\frac{1}{2}|\partial\chi_J|^2-U_{\text{eff}}(\chi_J,\xi)\right]~,\text{where}~~~U_{\text{eff}}(\chi_J,\xi)=\frac{V_{\text{eff}}}{\Omega^4}~.
\end{equation}
Here $\hat{R}$ is the Ricci scalar computed using the rescaled metric $\hat{g}_{\mu\nu}$ and $U_{\text{eff}}(\chi_J,\xi)$ is explicitly computed in Appendix.~\ref{AHiggsInfAppendix}. Now, after obtaining the effective action, we set the fields on-shell by removing the background current $J=0$. Assuming a flat FRW spacetime, the Friedmann equation and the equation of motion for the inflaton (Higgs singlet after redefinition) are
\begin{eqnarray}\label{InfEOM}
	&&3M_p^2H^2=\frac{1}{2}\dot{\chi}_0^2+U_{\text{eff}}(\chi_0,\xi)\\
	&&\ddot{\chi}_0+3H\dot{\chi}_0+\frac{\partial U_{\text{eff}}(\chi_0,\xi)}{\partial \chi_0}=0~.
\end{eqnarray}
This is just the normal pair of equations that governs the background dynamics during inflation. Usually the potential is flat such that the inflaton is drawn into the slow-roll attractor solution. The potential gradient term is now explicitly dependent on the gauge parameter $\xi$. This fact inevitably leads to the consequence that the slow roll parameters receive a gauge-dependent quantum correction:
\begin{eqnarray}
	\epsilon&\equiv&-\frac{\dot{H}}{H^2}\approx\epsilon_U=\frac{M_p^2}{2}\left(\frac{\partial_{\chi_0}U_{\text{eff}}(\chi_0,\xi)}{U_{\text{eff}}(\chi_0,\xi)}\right)^2\\
	\eta&\equiv&\frac{\dot{\epsilon}}{H\epsilon}\approx 4\epsilon_U-2\eta_U=2M_p^2\left[\left(\frac{\partial_{\chi_0}U_{\text{eff}}(\chi_0,\xi)}{U_{\text{eff}}(\chi_0,\xi)}\right)^2+\frac{\partial_{\chi_0}^2U_{\text{eff}}(\chi_0,\xi)}{U_{\text{eff}}(\chi_0,\xi)}\right]~.
\end{eqnarray}
Furthermore, the e-folding number also becomes gauge dependent:
\begin{equation}
	N_\text{CMB}=\int^{t_\text{end}}_{t_\text{CMB}} H dt\approx\int_{\chi_\text{CMB}}^{\chi_\text{end}}\frac{d\chi_0}{\sqrt{2\epsilon_U(\chi_0,\xi)}}~,
\end{equation}
where $\chi_\text{CMB}$ and $\chi_\text{end}$ are respectively the field values of the on-shell inflaton solution $\chi_0(t)$ at the time when the CMB scale exits the horizon and at the time when inflation is ended. In order to examine the impact of this apparent gauge dependence on direct inflationary observables, we can consider the scalar spectral index and tensor-to-scalar ratio,
\begin{equation}
	n_s-1\equiv \frac{d\ln P_\zeta}{d\ln k}\approx -2\epsilon-\eta,~~~r\equiv \frac{P_\gamma}{P_\zeta}\approx 16\epsilon~,
\end{equation}
where $P_\zeta$ and $P_\gamma$ are the scalar and tensor power spectra. 
\begin{figure}[h!]
	\centering
	\includegraphics[width=7.2cm]{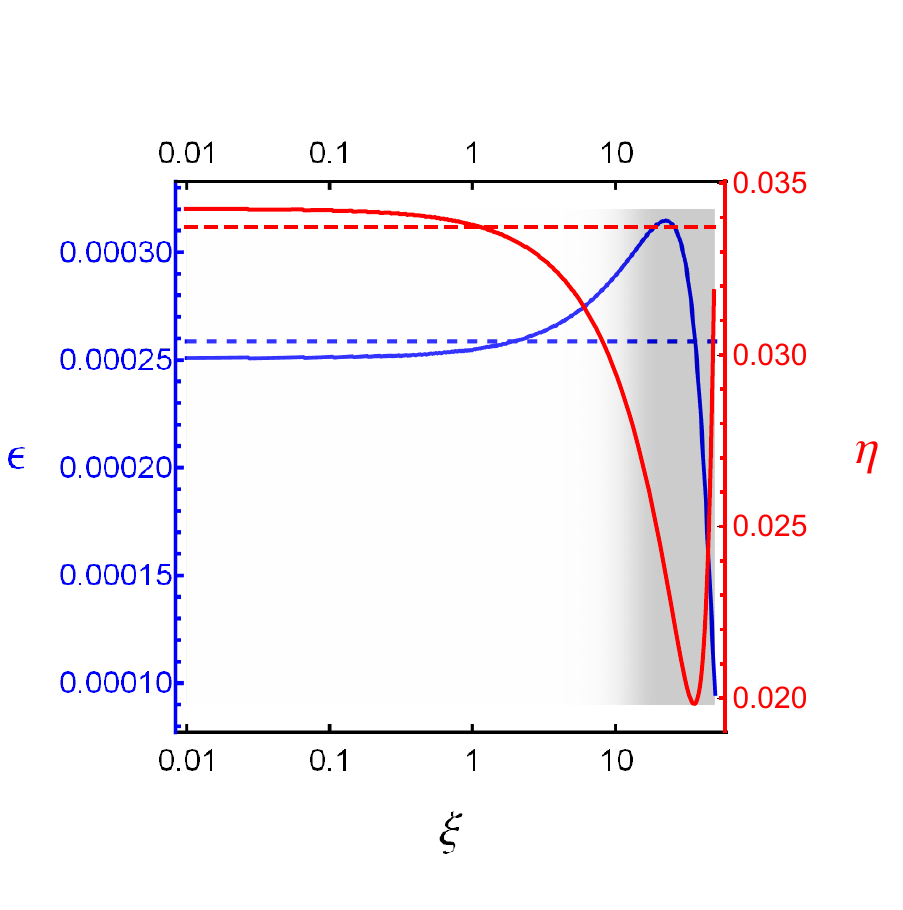}~~~~~~\includegraphics[width=9.5cm]{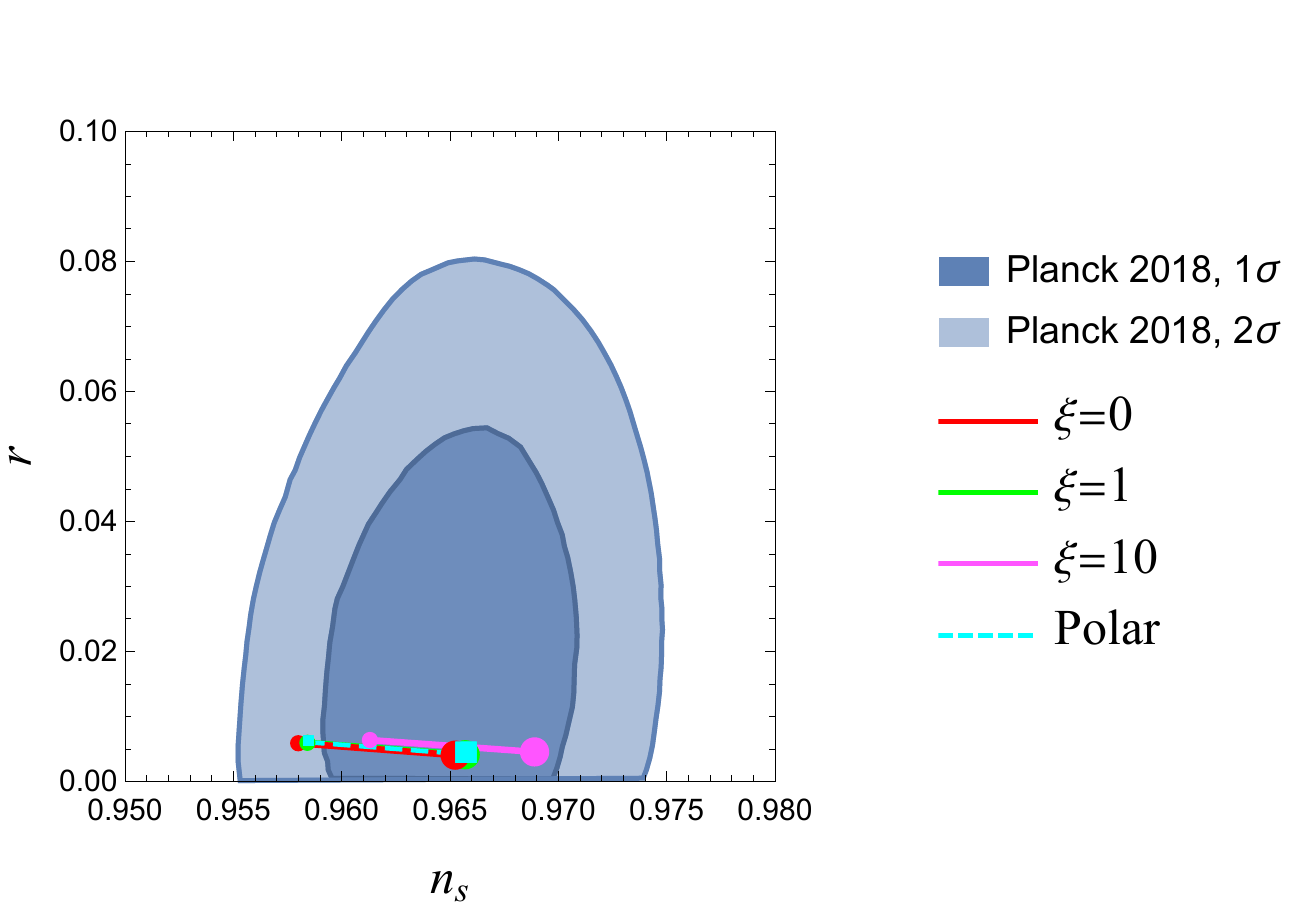}
	\caption{Left panel: The $\xi$ dependence of the slow-roll parameters at one-loop level for different field coordinates. The $\xi$-dependent solid lines are computed in conventional Cartesian coordinates while the $\xi$-independent dashed lines are computed in the polar field space coordinates. The gray region  is where $\xi\gtrsim 4\pi$ and gauge dependence threatens loop expansion validity and should not be taken seriously. Right panel: The apparent gauge dependence of the $n_s$-$r$ prediction in the Higgs inflation model at one-loop level. The smaller points correspond to $N_{\text{CMB}}=50$ while the larger points correspond to $N_{\text{CMB}}=60$. The dashed cyan line represents the gauge-invariant $n_s$-$r$ prediction in our proposed polar field coordinates. The smaller cyan square corresponds to $N_{\text{CMB}}=50$ while the larger cyan square corresponds to $N_{\text{CMB}}=60$. The contours are from TT,TE,EE+lowE+lensing+BK14 constraints with Planck 2018 data. The other parameters are chosen to be $\lambda=0.13, e=0.3, m=0, \alpha=17000,\mu/M_p=0.049$ to have a viable inflation model. See more details in Appendix ~\ref{AHiggsInfAppendix}.}\label{ns-r-xiDep}
\end{figure}

Using the effective action computed in the Cartesian field coordinates in the $R_\xi$ gauge, we numerically solve the full EOM (\ref{InfEOM}) and plot the $\xi$-dependence of the inflationary observables in Fig.~\ref{ns-r-xiDep}. The details of calculation is given in Appendix ~\ref{AHiggsInfAppendix}. As shown in the figure, within the valid range of one-loop result, the impact of apparent gauge dependence on inflationary observables is \textit{not} insignificant. In fact, this $\xi$-dependence is partly degenerate with other parameters in the model, for example, e-folding numbers. 

We point out that this gauge dependence problem is not only subjected to the particular Abelian Higgs inflation we mention here. It is generically present in any inflation models where the full inflaton field couples to gauge fields, giving a non-negligible uncertainty to the prediction of inflationary observables. Therefore in order to achieve more accurate constrains on the physical parameters in future CMB precision measurements, the uncertainty due to unphysical $\xi$ choices must be resolved.

\section{A wise choice of fundamental fields}\label{PolarExamps}
The apparent gauge dependence of the inflationary observables can be traced to several possible origins. 

Working backwards the derivation in the previous section, we see that the $\xi$ dependence in the final observables first appears in the potential-shape slow-roll parameters $\epsilon_U ,\eta_U$ which are approximations to the original ones $\epsilon,\eta$, up to errors $\mathcal{O}(\epsilon_U^2,\eta_U^2)$. One might question the possibility that higher order corrections will render $\epsilon,\eta$ gauge-invariant. However, numerical solution in Fig.~\ref{ns-r-xiDep} shows that with $\mathcal{O}(\epsilon_U^2,\eta_U^2)$ corrections included, the full $\epsilon,\eta$ parameters still depend on $\xi$.

Therefore the equations of motion (\ref{InfEOM}) already contain unphysical information. Apparently this is because the inflaton potential $U_{\text{eff}}(\chi_0,\xi)$ in the Einstein frame depends on $\xi$ explicitly, which is not washed out by the procedure of looking into observables. Since (\ref{InfEOM}) follows from the variational principle $\frac{\delta S_{\text{eff}}}{\delta \chi_J}=-J=0$, the removal of the background current should guarantee an equation of motion that gives out gauge-independent final observables. Its failure suggests that we have used the wrong effective action (\ref{EinEffAction}). As field redefinitions do not change equations of motion, the effective action in Jordan frame (\ref{JorEffAction}) is also problematic. Thus viewed in this way, one possible cause for the gauge dependence problem is the procedure of expanding the originally non-local effective action and truncating to second order in field gradients. When the inflaton background is set free to roll, the whole tower of field gradients may contribute to the rolling solution. 

This is the case for the scenario without gravity, where $\phi_0$ quickly rolls off a steep potential. When the rolling speed is too fast, $i.e.$, $|\partial\phi_0|\gtrsim \partial_\phi^2 V$, the adiabatic condition for perturbations will be violated and real particles will be produced. The system cannot be modeled by a uniform background field $\phi_0(t)$ anymore, and the quantum corrections in the average field value $\phi_0(t)$ receive non-local contribution from the real field quanta. To exactly evaluate the quantum corrections and check gauge dependence, one may need a non-perturbative treatment (such as that proposed recently in \cite{Pimentel:2019otp}). 

However, in the scenario with gravity, obviously the Hubble friction slows down the rolling speed. The system evolves adiabatically and the non-local contributions are exponentially suppressed. In this case, the truncation procedure seems valid since the field gradients are indeed small. The gauge dependence issue then poses a question mark on our neglect of spacetime curvature in the beginning (\ref{gToEta}). Then one may need to take the expansion of spacetime into account and use in-in formalism in a quasi-de Sitter spacetime to re-evaluate the loop diagrams that give rise to the effective potential. There are some preliminary results on this possibility. For example, it is shown in \cite{George:2012xs} that the value of the effective action for a rolling inflaton is $\xi$-independent on-shell, which is in agreement with the Nielsen identity. Yet to explicitly demonstrate the gauge independence of the whole set of inflationary observables, there is still a long way to go.

In either cases, the computational work needed to resolve the gauge dependence issue is huge, if not impossible. As a result, we propose a third remedy which is much easier and more practical.

\begin{figure}[h!]
	\centering
	\includegraphics[width=8cm]{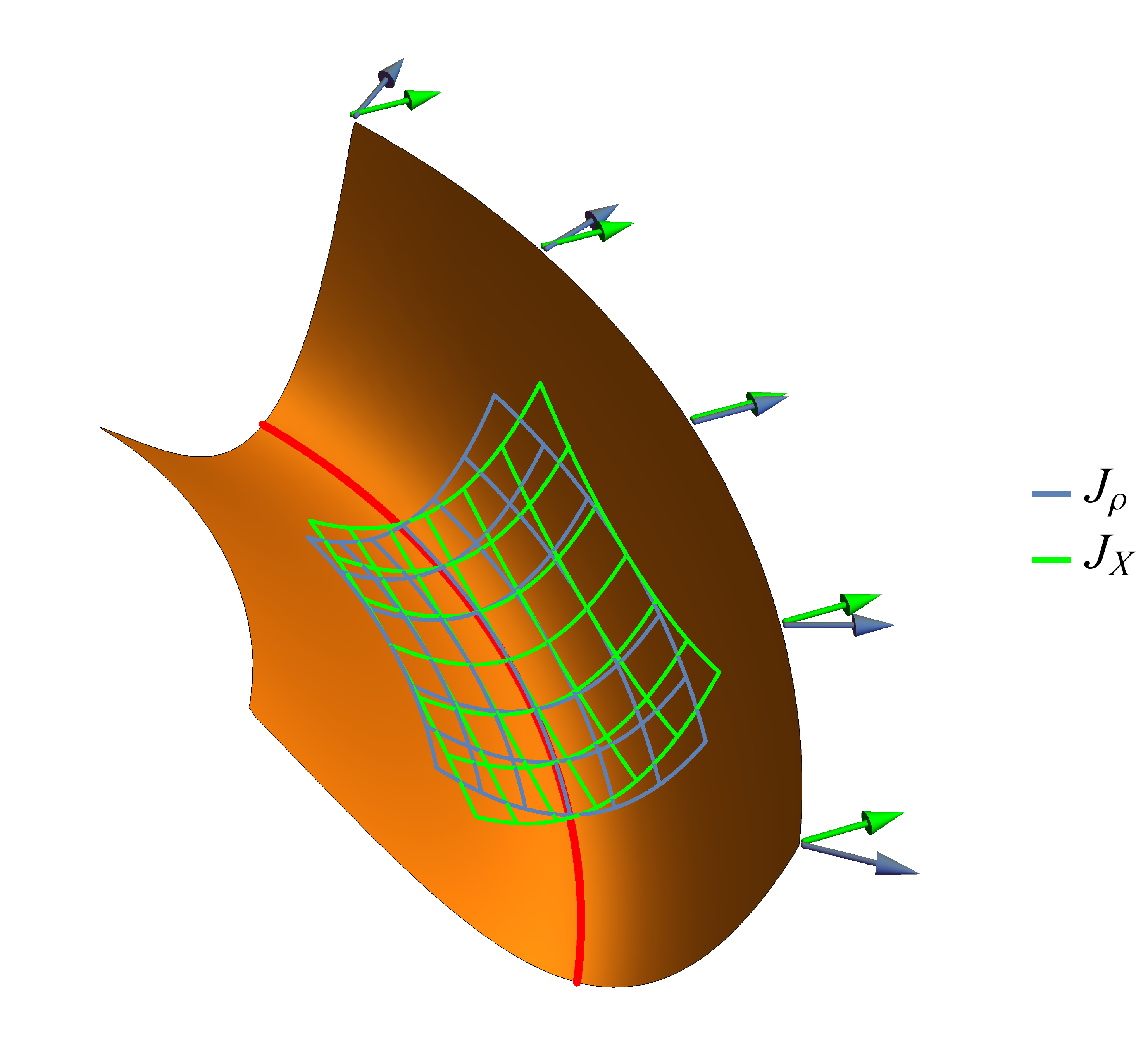}\includegraphics[width=9.5cm]{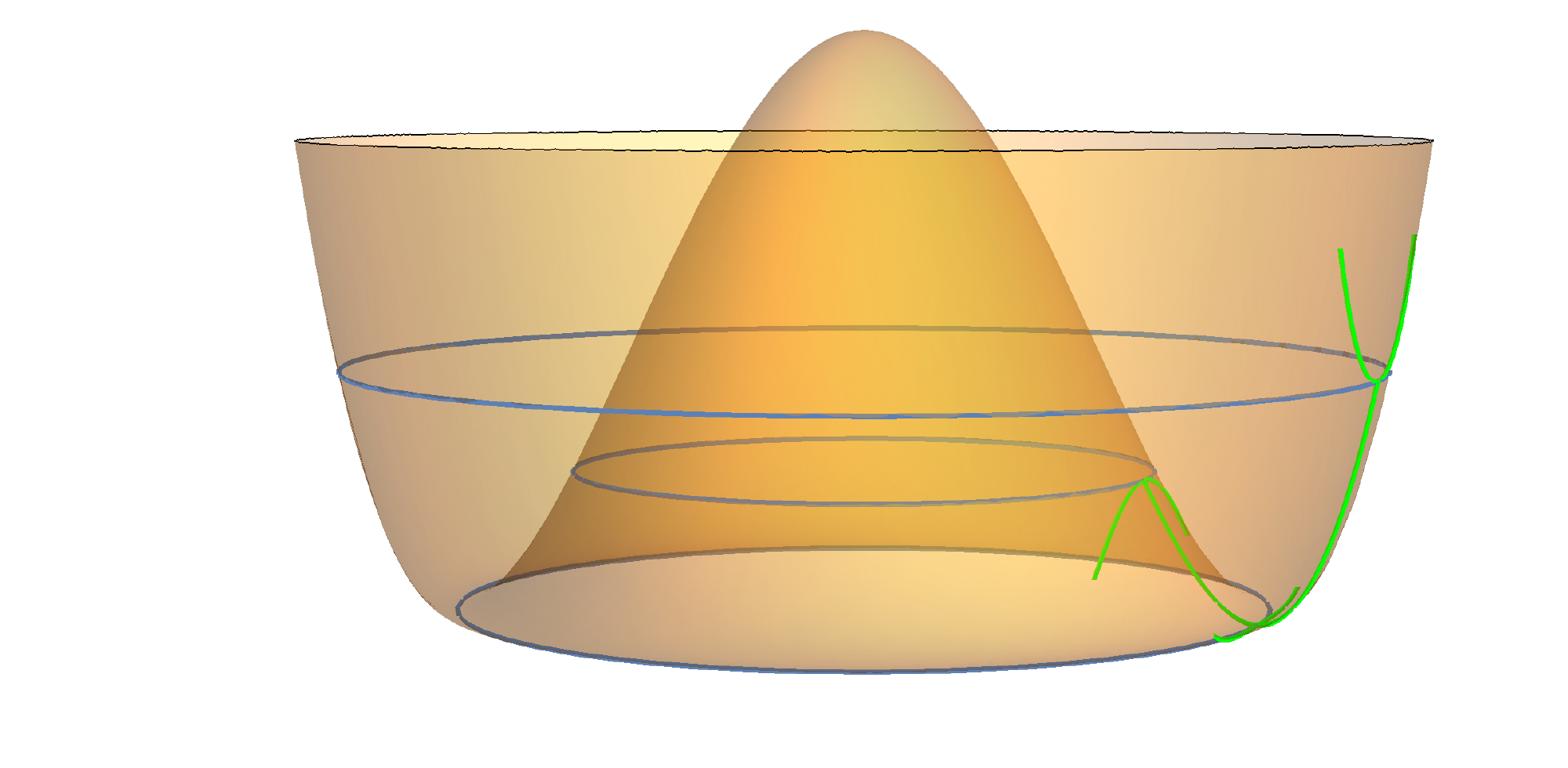}
	\caption{Left panel: A wise coordinate choice induces a symmetric background current. The green lines correspond to Cartesian coordinates while blue lines correspond to polar coordinates. In turn, $J_X$ breaks $U(1)$ while $J_\rho$ preserves it. Right panel: In polar fields, the Goldstone direction is flat thus massless. In contrast, in Cartesian fields, the Goldstone is generally massive away from the minimum.}\label{Cartesian Vs Polar}
\end{figure}
In the literature, Cartesian coordinate fields are usually used to construct the effective potential. The physical meaning for the effective potential is then the minimal energy density of the system under the constraint of an average field value $\bar{\phi}$ \cite{Coleman:1973jx}. To impose the constraint, one needs to apply an auxiliary background current $J$ suited to the coordinate system. Therefore, if the coordinate system does not respect the gauge symmetry, its corresponding current naturally breaks gauge invariance and leads to a gauge-dependent effective potential off-shell. For the purpose of determining the true vacuum, this is convenient since the vacuum lies at a field configuration without the support of a background current. The same is true if one considers the on-shell scattering amplitudes near the vacuum. As a result, it is stated that effective potential is only physical at its local extrema, where fields are on-shell\cite{Patel:2011th,Andreassen:2014eha}. However, as we have seen above, there is no practical solution to the sensitivity of inflationary observables to shape of the effective potential \textit{away} from the local extrema.

Thus the consideration of cosmological applications leads us to the \textit{off-shell} gauge-invariant effective potential. We explore the direct use of a gauge-invariant auxiliary current, naturally associated with polar-like coordinates (see Fig.~\ref{Cartesian Vs Polar}) that catch the symmetry of the theory, as suggested in \cite{Tye:1996au}. We use a $J_i \varphi^i$ source term for which $J_i \delta_{\scaleto{\rm  BRST}{3pt}} \varphi^i = 0$. This guarantees a gauge-invariant effective action. This can be clearly observed in an example studied in \cite{Tye:1996au}, where the Abelian Higgs effective potential is computed using the polar expression of the scalar field. There, the gauge dependence cancels out and the resulting effective potential is clearly gauge-independent. This eventually leads to a clean and invariant effective potential which is simple to compute. Moreover, the resulting effective potential exactly matches the one computed in the unitary gauge, thus in agreement with the conclusions of \cite{Dolan:1974gu}. Here, we will go further to check whether this prescription works for the cases of an $SU(2)$ theory in the Fundamental representation, $SU(2)$ in the adjoint representation, and $SU(2) \times U(1) $ (the Standard model). For generality, we compute two of them in the covariant gauge and the other two in $R_\xi$ gauge. We will explicitly compute the gauge dependent part of the effective potential, and show that it vanishes in the case of a symmetric current term.

\subsection{The simplest example, U(1)}\label{U1Examp}
For pedagogical reasons, we will rederive first the Abelian Higgs one-loop effective potential, with polar field space coordinates as the fundamental fields. Coupling the polar fields with the corresponding currents and choosing the covariant gauge family, the Lagrangian reads
\begin{equation}
\mathcal L = | D \phi|^2 -V(|\phi|) -\frac{1}{4} F^2 - \frac{1}{2 \xi} (\partial A)^2- \bar c ~ \partial^2 c +  J_\rho \rho + J_\theta \theta~. 
\end{equation}
Setting $\phi= (\bar \phi + \rho) \, e^{i \theta} $, then, by applying the steepest decent method (which forces $J_\theta$ to 0), we get the relevant quadratic Lagrangian
\begin{eqnarray} \nonumber  \mathcal L_2 &=&-\rho \left(\partial^2+6\lambda\bar{\phi}^2-m^2\right)\rho \\
\nonumber&&- {\bar \phi}^2 \theta \partial^2\theta + 2\, e \,\bar \phi A_\mu \partial^\mu \theta-\bar c ~ \partial^2 c\\
&&+\frac{1}{2}A_\mu\left[(\partial^2+2e^2\bar{\phi}^2)\eta^{\mu\nu}-\left(1-\frac{1}{\xi}\right)\partial^\mu\partial^\nu\right] A_\nu  ~, 
\end{eqnarray}
where we have denoted $\bar \phi \equiv \langle \phi \rangle_{_J}$ as the constrained radial field value. Then, after Legendre transformation and integrating over perturbations, the effective potential becomes a trace over the Hessian matrix of the constrained Lagrangian at $\phi=\bar{\phi}$,
\begin{eqnarray}
\nonumber V_{\rm eff}&=&\frac{i}{2} \Tr \ln \left[-\left( \partial^2 \eta^{\mu \nu} +(1 - \frac{1}{\xi} ) \partial^{\mu} \partial^{\nu} +2 e^2 \bar \phi^2 \eta^{\mu \nu}\right) \right] \\
&&+ \frac{i}{2}  \Tr \ln \left[ - \partial^2  + \frac{2 e^2 \bar \phi^2 \partial^\mu \partial^\nu}{\partial^2 + 2 e^2 \bar \phi^2 } \left(\eta_{\mu \nu} + \frac{\partial_\mu \partial_\nu(1-\xi)}{-\partial^2 - 2 \xi e^2 \bar \phi^2 } \right)  \right] \nonumber  \\
 &&+  \rm gauge \, \, independent  \, \,terms \\
 \nonumber \\
\nonumber &=&  \frac{ \Gamma(\frac{-d}{2})} { (4 \pi)^ \frac{d}{2}} 3 (2 e^2 \bar{\phi} ^2)^\frac{d}{2} +\frac{ \Gamma(\frac{-d}{2})} { (4 \pi)^ \frac{d}{2}} (2 \xi e^2 \bar{\phi} ^2)^\frac{d}{2}\\
\nonumber && +  \frac{i}{2} \int \frac{d^dk}{(2 \pi)^d} \ln \left[  -k^4 \right] - \frac{ \Gamma(\frac{-d}{2})} { (4 \pi)^ \frac{d}{2}} (2\xi e^2 \bar \phi ^2 )^\frac{d}{2}  \\
&& + {\rm gauge \, \, independent  \, \,terms}~.  
\end{eqnarray}
The first two terms come from the gauge fields, the last two come from the Goldstone. 
Here we have only spelled out the terms with explicit $\xi$ dependence and omitted the rest. As is clear, the gauge dependence cancels out everywhere. In the theory with Cartesian coordinates as fundamental fields, this happens at the vacuum, anywhere else the gauge dependence is explicit.\\
This is a clear example demonstrating how the right choice of the fundamental fields (and consequently, the source term), can resolve the off-shell gauge dependence of the effective potential. On the other hand, if we choose a current that breaks a symmetry, then we should not be surprised that that symmetry is broken at the level of the effective potential or effective action. In this formalism, one can easily compute an effective action which is gauge independent on-shell and off-shell, from which the derived physical quantities will all be gauge independent.

\subsection{$SU(2)$ in fundamental representation}
Now we move on to explore this prescription in more complicated settings. First, we start with an $SU(2)$ model with the scalar transforming under the fundamental representation. Upon spontaneous symmetry breaking, all three generators of $SU(2)$ are broken by the scalar VEV. Then, the scalar field can be parametrized as 
\begin{eqnarray}   \phi &=&  e ^{i \theta^a \tau^a}
\begin{pmatrix}
0 &\\
\bar \phi + \rho&
\end{pmatrix}, \nonumber \\
 \tau^a  &=& \frac{1}{2}\sigma^a \, \, \, ; \, \, \, \, j =1, 2 ,3  ~,
\end{eqnarray}
with $\rho$ playing the role of the singlet radial field and $\theta^a$ being the angular fields transforming non-linearly under $SU(2)$.
Then we turn on the gauge fields by a minimal coupling $D_\mu\phi=(\partial_\mu-ig A_\mu^a\tau^a)\phi$, where
\begin{eqnarray}
        A_\mu &=& A^a _\mu \tau^a  \, \, \, ; \, \, \, a=1, 2, 3. \nonumber  \\
          F_{\mu \nu} &=& \partial_\mu A_\nu - \partial_\nu A_\mu - ig [A_\mu, A_\nu]~. 
\end{eqnarray} 
In the covariant gauge $\frac{1}{2\xi} (\partial A )^2$,  after sufficient manipulation, the relevant quadratic Lagrangian becomes
\begin{eqnarray}
\nonumber\mathcal L_2 &=& - \rho  \left(\partial^2\ + 6 \lambda \bar \phi ^2  - m^2 \right)\rho\\
\nonumber&&-\frac{\bar \phi ^2}{4} \theta^a\partial^2 \theta^a-\frac{1}{2}g\bar{\phi}^2 A_\mu^a \partial^\mu\theta^a  - \bar c^a ~\partial^2 c^a  \\
&&+\frac{1}{2} A^a_\mu \left[\eta^{\mu \nu}\left( \partial^2 + \frac{1}{2}g^2\bar{\phi}^2\right) - \left(1 - \frac{1}{\xi} \right)\partial^\mu \partial^\nu\right]A_\nu^a ~.
\end{eqnarray}
This gives a one-loop effective potential which reads
\begin{eqnarray}
 \nonumber V_{\rm eff}&=&\Tr \ln  \left[ k^2 - \frac{ \frac{1}{2} g^2 \bar \phi^2  k^2 }{-k^2 + \frac{1}{2} g^2 \bar \phi^2 }\left(1 - \frac{ (1-\xi) k^2}{k^2 - \frac{1}{2}  \xi  g^2 \bar \phi^2} \right) \right]\\
 \nonumber&& + \Tr \ln \left[ (-k^2 + \frac{1}{2} g^2 \bar \phi^2)^3 \, ( -\frac{k^2}{\xi} + \frac{1}{2}  g^2 \bar \phi^2)  \right]  \\
 \nonumber 
&&+ \,{\rm gauge  \, \, independent  \, \, terms.} 
\end{eqnarray}
Which in turn boils down to
\begin{eqnarray} &&\frac{ \Gamma(\frac{-d}{2})} { (4 \pi)^ \frac{d}{2}}3\left( \frac{1}{2} g^2 \bar \phi ^2  \right)^\frac{d}{2}  -\frac{ \Gamma(\frac{-d}{2})} { (4 \pi)^ \frac{d}{2}}  \left(\frac{1}{2}   \xi g^2 \bar \phi^2 \right)^\frac{d}{2}  + \frac{ \Gamma(\frac{-d}{2})} { (4 \pi)^ \frac{d}{2}} \left( \frac{1}{2}  \xi g^2 \bar \phi^2 \right)^\frac{d}{2} \nonumber  \\
&&  + \,{\rm gauge  \, \, independent  \, \, terms.}
\end{eqnarray}
Where the first two terms come from the gauge fields and the last one represents the Goldstones. 
Finally, after cancellation, we are left solely with gauge independent terms. 

\subsection{$SU(2)$ in the adjoint representation}
We can also work out the $SU(2)$ theory in the adjoint representation, $i.e.$ ,the Georgi-Glashow model which was originally introduced in order to explain the boson mass. 
After the adjoint scalar acquires a VEV, two of the $SU(2)$ generators are broken and only one generator preserves the scalar VEV, which lies on a unit sphere $S^2\cong SU(2)/U(1)$. Choosing the polar coordinate is equivalent to choosing a north pole on this coset manifold $S^2$. Without loss of generality, we will choose the direction of the scalar VEV as the north pole, which is always possible by applying suitable similarity transformations. In its vector presentation, this theory consists of an adjoint scalar field which after a suitable similarity transformation reads
\begin{eqnarray} \phi =  e ^{i \alpha^j \tau^j} \begin{pmatrix}
0 &\\
0 &\\
\bar \phi + \rho&
\end{pmatrix},  \, \, \, \, \,
j=1,2~.
\end{eqnarray}
The gauge fields are coupled minimally with $\phi$ through a covariant derivative 
\begin{eqnarray}  
\left[ D_\mu \phi \right]^a = \partial_\mu \phi^a +  g \epsilon^{abc} A_\mu ^b \phi^c. 
\end{eqnarray}
This time, we use the $R_\xi$ gauge $\frac{1}{2\xi} (\partial A - 2 g \xi \bar \phi^2 \alpha )^2$ where the mixing between gauge bosons and Goldstones is canceled. We focus on the relevant quadratic Lagrangian,
\begin{eqnarray}
\mathcal L _2  &=& -\rho (\partial^2  +6 \lambda \bar \phi ^2- m^2 ) \rho \nonumber \\ 
 &&- \bar \phi^2 \alpha^j (\partial^2 + 2\xi g^2\bar \phi^2  ) \alpha^j \nonumber \\
 &&+\frac{1}{2} A^j_\mu \left[\eta^{\mu \nu}\left( \partial^2 + 2g^2\bar{\phi}^2\right) - \left(1 - \frac{1}{\xi} \right)\partial^\mu \partial^\nu\right]A_\nu^j \nonumber \\ 
&&- \bar c^j ( \partial^2 + 2\xi g^2 \bar \phi^2 ) c^j~.
\end{eqnarray}
Notice that here we have only written down the terms relevant to the computation of the effective potential. The unbroken $U(1)$ part can be gauge-fixed independently (for example, with a different gauge parameter $\xi'$) and will not influence the resulting effective potential $V_{\text{eff}}(\bar{\phi},\xi)$. Then after integration and Legendre transformation it gives,
\begin{eqnarray} \frac{ \Gamma(\frac{-d}{2})} {2 (4 \pi)^ \frac{d}{2}}  \left[ 3 (2 g^2 \bar \phi^2) ^\frac{d}{2} + (2 \xi g^2 \bar \phi^2 ) ^\frac{d}{2} \right] + \frac{ \Gamma(\frac{-d}{2})} {2 (4 \pi)^ \frac{d}{2}} (2 \xi g^2 \bar \phi^2 ) ^\frac{d}{2} - \frac{ \Gamma(\frac{-d}{2})} { (4 \pi)^ \frac{d}{2}} (2 \xi g^2 \bar \phi^2) ^\frac{d}{2}~.
\end{eqnarray}
Here, the first two terms are due to the gauge fields, the third to the Goldstones and the last is the ghost contribution.
Clearly, the gauge dependence cancels out, just as expected.

\subsection{The Standard Model}
Here we work out the effective potential for the Glashow-Salam-Weinberg (GSW) model without fermions, choosing polar fields as the fundamental fields. 
There are 4 generators for $SU(2) \cross U(1)$. When the symmetry is broken to $U(1)$, three generators are broken and one is left. 
However, as is known in the GSW model, the unbroken generator is a linear combination of the $U(1)$ and one of the $SU(2)$ generators. 
After a suitable similarity transformation, the scalar field in the polar-like coordinates can be written as 
\begin{equation}
\phi = e ^{ i \left( \alpha^+ T^+ + \alpha^- T^- \right) - i \alpha_Z T_Z } \begin{pmatrix}
0 &\\
\bar \phi + \rho&
\end{pmatrix}~,
\end{equation}
where we have denoted $T^\pm=\tau^1\pm i\tau^2$, $T_Z=\tau^0-\tau^3$ and $T_A=\tau^0+\tau^3$.

Turing on the minimal coupling with gauge fields, the scalar covariant derivative reads
\begin{equation}
D_\mu \phi= \partial_\mu \phi- \!\frac{ig}{\sqrt{2}}(W^+_\mu T^+\! +\! W^-_\mu T^-\!)\phi-\frac{ig}{c_w}Z_\mu\left[\left(\frac{1}{2}-s_w^2\right)T_A-\frac{1}{2}T_Z\right]\phi-ig s_w A_\mu T_A\phi~,
\end{equation}
where, as usual, $s_w=\frac{g'}{\sqrt{g^2+g'^2}}$ and $c_w=\frac{g}{\sqrt{g^2+g'^2}}$. After expansion to the quadratic order, we get the scalar kinetic term as
\begin{eqnarray}  
\nonumber\|D\phi\|^2 &=& (\partial \rho) ^2 + \bar \phi ^2 \partial \alpha ^+ \partial \alpha ^- + \bar \phi ^2 (\partial \alpha_Z)^2 \\
\nonumber&&+ \frac{1}{2}g^2\bar{\phi}^2 W^+ W^- + \frac{1}{4c_w^2}g^2\bar{\phi}^2 Z^2\\
&&-\frac{g\bar{\phi}^2}{\sqrt{2}}\left(W^+\partial\alpha^- + W^-\partial\alpha^+\right)-\frac{g\bar{\phi}^2}{c_w}Z\partial\alpha_Z~. 
\end{eqnarray}
Now we fix the gauge into the $R_\xi$ family in order to cancel the mixed terms. Let the gauge fixing term be 
\begin{equation}
G^2 =-\frac{1}{2 \xi} ( \partial W^+ +\sqrt 2 \xi g\bar \phi ^2 \alpha^+)( \partial W^- +\sqrt 2 \xi g\bar \phi ^2 \alpha^-) - \frac{1}{2 \zeta} \left(\partial Z + \frac{\zeta  g \bar \phi^2}{c_w} \alpha _Z\right)^2~. 
\end{equation}
Then, the relevant gauge-fixed quadratic Lagrangian reads
\begin{eqnarray} 
\nonumber\mathcal L_2 &=&-\rho\left(\partial^2+6\lambda\bar{\phi}^2-m^2\right)\rho\\
\nonumber&&-\bar{\phi}^2\alpha^+\left(\partial^2+\xi g^2\bar{\phi}^2\right)\alpha^-\\
\nonumber&&-\bar{\phi}^2\alpha_Z\left(\partial^2+\frac{\zeta}{2c_w^2}g^2\bar{\phi}^2\right)\alpha_Z\\
\nonumber&&+\frac{1}{2} W_\mu^+ \left[\eta^{\mu \nu}\left( \partial^2 + g^2\bar{\phi}^2\right) - \left(1 - \frac{1}{\xi} \right)\partial^\mu \partial^\nu\right]W_\nu^-\\
\nonumber&&+\frac{1}{2} Z_\mu \left[\eta^{\mu \nu}\left( \partial^2 + \frac{1}{2c_w^2}g^2\bar{\phi}^2\right) - \left(1 - \frac{1}{\zeta} \right)\partial^\mu \partial^\nu\right]Z_\nu\\
\nonumber&&-\bar{c}^+\left(\partial^2+\xi g^2\bar{\phi}^2\right)c^-\\
&&-\bar{c}_Z\left(\partial^2+\frac{\zeta}{2c_w^2}g^2\bar{\phi}^2\right)c_Z~.
\end{eqnarray}

As a result, the gauge-dependent terms are
\begin{eqnarray}  && \frac{ \Gamma(\frac{-d}{2})} { (4 \pi)^ \frac{d}{2}} 2 (\xi g^2 \bar \phi^2) ^\frac{d}{2} + \frac{ \Gamma(\frac{-d}{2})} { (4 \pi)^ \frac{d}{2}} 2 (\xi g^2 \bar \phi^2) ^\frac{d}{2} - \frac{ \Gamma(\frac{-d}{2})} { (4 \pi)^ \frac{d}{2}} 2 \times 2 (\xi g^2 \bar \phi^2) ^\frac{d}{2}  \nonumber  \\
&& \frac{ \Gamma(\frac{-d}{2})} { (4 \pi)^ \frac{d}{2}} \left(\frac{\zeta}{2c_w^2}g^2\bar{\phi}^2\right) ^\frac{d}{2} +  \frac{ \Gamma(\frac{-d}{2})} { (4 \pi)^ \frac{d}{2}} \left(\frac{\zeta}{2c_w^2}g^2\bar{\phi}^2\right) ^\frac{d}{2} - \frac{ \Gamma(\frac{-d}{2})} { (4 \pi)^ \frac{d}{2}} 2  \left(\frac{\zeta}{2c_w^2}g^2\bar{\phi}^2\right) ^\frac{d}{2}~.
\end{eqnarray}
In each line the first term refers to the gauge fields, the second to the Goldstones, and the third to the ghosts. The first line comes from the $W^\pm$'s and their related fields while the second is contributed by $Z$ and its related fields. Thus we see again that they cancel each other and the effective potential is explicitly gauge-independent.

In summary, adopting polar-like scalar coordinates amounts to an application of symmetric background currents which preserves the gauge-invariance of the final effective potential. This, in turn, gives rise to a gauge-independent prediction of inflationary observables. For the Abelian Higgs inflation model in Sect.~\ref{Cosmo}, we numerically solve the inflaton dynamics and plot our prediction of $n_s$ and $r$ for polar coordinate choice in Fig.~\ref{ns-r-xiDep}. The result for polar coordinate choice differs from that of the whole $\xi<\infty$ gauge family for traditional Cartesian coordinates
. In Sect.~\ref{unitaryLim}, we will argue that the polar coordinate result actually matches that of the unitary gauge, namely $\xi\to \infty$ with the cutoff held fixed.

Before we conclude this section, we point out that in the literature, there exist other ways of obtaining off-shell gauge-independent effective potential. For example, \cite{Boyanovsky:1996dc} computes the effective potential in the Hamiltonian formalism using a gauge-invariant order parameter. However, their result is different from ours. Therefore, which result is more accurate must be left for more rigorous calculations or experiments in the future.

\section{Gauge dependence identities}\label{GaugeDepId}

In this section we first review the derivation of the gauge dependence identities \cite{Nielsen:1975fs} using the functional notation introduced by \cite{DeWitt:1967ub,Vilkovisky:1984st}, as presented in \cite{Kobes:1990dc, Carrington:2003ut}. We emphasize along the derivation that the asymmetric current term is the source of all gauge dependence appearing in the effective action. We then discuss what seems to be the most reasonable way to achieve its off-shell gauge invariance non-perturbatively, namely by using a current coupled to the gauge-singlet radial field in the polar coordinates on the scalar manifold.

We start by defining $\varphi_i$ to span all the fields in a given theory\footnote{Notice that this DeWitt notation will only be used in this section to avoid cluttering irrelevant symbols.}. The $i$ index accounts for spacetime variables, Lorentz index, and the group index, altogether (e.g. $\varphi^k = A_\mu ^a(x)$). On the other hand, a Greek index (e.g. $\mathcal F_\alpha \equiv \partial _\mu A^\mu _a (x) $) accounts only for spacetime and group indices. The gauge transformation is defined by $\delta_g\varphi_i=\theta^\alpha\delta_\alpha\varphi_i$, where $\theta^\alpha$ is a transformation parameter and $\delta_\alpha$ is the gauge generators defined at a local spacetime point. The gauge algebra is $[\delta_\alpha,\delta_\beta]=f^{\gamma}_{~\alpha\beta}\delta_\gamma$, where $f^{\gamma}_{~\alpha\beta}$ consists of ordinary structure constants as well as derivatives of the spacetime Dirac delta function \cite{Polchinski:1998rq}. We will call $\mathcal F^A$ the gauge fixing function, with $\{A\}\subset\{\alpha\}$. Now, the argument goes as follows.

We start with the gauge-fixed partition function 
\begin{eqnarray}
	Z[J]=\int \mathcal{D} \varphi \,  \mathcal{D} \eta \,\mathcal{D}\bar \eta\, e^{i\left(S -  \frac{1}{2\xi}\mathcal  F_A \mathcal F^A  + \,  \bar \eta_A \delta_\alpha \mathcal{F}^A \eta^\alpha \, +  J^i \varphi_i\right)}.
\end{eqnarray}
The ghost propagator is defined to be the inverse of the Faddeev-Popov operator $\delta_\alpha \mathcal{F}^A$, 
\begin{eqnarray}
	\delta_\alpha \mathcal{F}^A \mathcal{G}^{\alpha}_{~B} = - \delta^A_{~B}~. \label{KeepFeq}
\end{eqnarray}
The above action is invariant under the BRST transformation, 
\begin{equation}
	\delta_{\scaleto{\rm BRST}{3pt}} \varphi_i = \zeta\eta^\alpha\delta_\alpha\varphi_i, \, \, \, \delta _{\scaleto{\rm  BRST}{3pt}} \bar \eta_A =  \zeta \mathcal F_A/\xi , \, \, \, \delta_{\scaleto{\rm  BRST}{3pt}} \eta^\alpha = -\frac{1}{2}\zeta f^\alpha_{~\beta\gamma}\eta^\beta \eta^\gamma~,
\end{equation}
where $\zeta$ is a constant Grassmann number. Now, consider a correlation function 
\begin{eqnarray}
 \langle O(\varphi,\bar \eta, \eta) \rangle \equiv e^{-i W} \int \mathcal D \varphi \mathcal \mathcal D \eta D \bar \eta\, O \, e^{i\left(S -  \frac{1}{2\xi}\mathcal  F_A \mathcal F^A  + \,  \bar \eta_A \delta_\alpha \mathcal{F}^A \eta^\alpha \, +  J^i \varphi_i\right)}~.
\end{eqnarray}
An easy but useful observation is that this function vanishes for an operator with odd power of ghost fields, following from ghost number conservation. An example that turns out to be useful is 
\begin{equation}
	\langle \bar \eta_A (x) G^A(x) \rangle = 0~,
\end{equation}
where $G$ can be a function of any non-ghost fields. Then, one can apply a BRST transformation to this equation, to obtain another form of the same equality, so that it encodes the outcomes of the BRST symmetry. 
\begin{equation} 
\delta _{\scaleto{\rm  BRST}{3pt}} \langle \bar \eta_A G^A(x) \rangle= \langle \delta _{\scaleto{\rm  BRST}{3pt}} [ \bar \eta_A G^A(x) ]  + i \bar \eta_A G^A(x) J^i \delta _{\scaleto{\rm  BRST}{3pt}}  \varphi_i  \rangle = 0~,
\end{equation}
where the second term comes from the non-zero background current support. We can rewrite this as
\begin{equation}
\left\langle\mathcal{F}_A G^A(x)/\xi-\bar{\eta}_A\eta^\alpha\delta_\alpha G^A(x)\right\rangle=-iJ^i\left\langle\eta^\alpha\delta_\alpha\varphi_i\bar{\eta}_AG^A(x)\right\rangle~.  \label{key}
\end{equation}\\
The identity (\ref{key}) is of crucial importance, hence we will come back to it later. To derive the Nielsen equation, we need to check how the connected-diagram-generating functional $W[J]$ varies with a change of the gauge choice.

So, we consider a change of our gauge choice $\mathcal F \rightarrow \mathcal F' = \mathcal F + \Delta \mathcal F$.
Then, $W$ becomes 
\begin{eqnarray} 
W' &=& -i \ln \int \mathcal{D} \varphi \,  \mathcal{D} \eta \,\mathcal{D}\bar \eta\, e^{i\left(S -  \frac{1}{2\xi}\mathcal  F_A^\prime \mathcal F^{\prime A}  + \,  \bar \eta_A \delta_\alpha \mathcal{F}^{\prime A} \eta^\alpha \, +  J^i \varphi_i\right)}\\
&=& -i \ln \int \mathcal{D} \varphi \,  \mathcal{D} \eta \,\mathcal{D}\bar \eta\, e^{i\left(S -  \frac{1}{2\xi}\mathcal  F_A \mathcal F^A  + \,  \bar \eta_A \delta_\alpha \mathcal{F}^A \eta^\alpha \, +  J^i \varphi_i\right)}\left(1-i\mathcal  F_A \Delta \mathcal F^A/\xi  + \,  i\bar{\eta}_A \delta_\alpha \Delta\mathcal{F}^A \eta^\alpha \right) \nonumber \\
&=& W - i\langle \mathcal{F}_A \Delta \mathcal F^A/\xi-\bar{\eta}_A \eta^\alpha\delta_\alpha \Delta\mathcal{F}^A  \rangle~,
\end{eqnarray}
with the change given by
\begin{equation}
	\Delta W = - i\langle \mathcal{F}_A \Delta \mathcal F^A/\xi-\bar{\eta}_A \eta^\alpha\delta_\alpha \Delta\mathcal{F}^A  \rangle~.\label{key2}
\end{equation}
Now we look at (\ref{key}), we find that for $G^A$ there identified with $\Delta \mathcal F^A$, it gives the RHS of (\ref{key2}) after integrating over $x$. Thus, identifying the two equations, we get
\begin{eqnarray} 
\Delta W = - J^i\left\langle\eta^\alpha\delta_\alpha\varphi_i\bar{\eta}_A\Delta\mathcal{F}^A\right\rangle~.\label{DeltaWeq}
\end{eqnarray}
Now, given that the effective action $\Gamma$ is the Legendre transform of $W$,
\begin{eqnarray}\label{Gamma} 
	\Gamma[\phi] = W[J] - J_i \varphi^i~,
\end{eqnarray} 
the variation in the effective action will be 
\begin{eqnarray} 
\Delta \Gamma = \frac{\delta \Gamma}{\delta \phi_i} \left\langle\eta^\alpha\delta_\alpha\varphi_i\bar{\eta}_A\Delta\mathcal{F}^A\right\rangle~,\label{key3}
\end{eqnarray}
yielding the Nielsen identity in a generalized form. Here, $\Delta \Gamma$ is the variation of the effective action under a change in of the gauge choice $\Delta \mathcal F$. As is clear, the RHS vanishes on-shell, since the $\frac{\delta \Gamma}{\delta \phi_i}$ term is equal to $J_i$ and thus vanishes. However, off-shell, it still survives. As explained in Sect.~\ref{PolarExamps}, there is nevertheless no easy solution to the apparent on-shell gauge dependence in the inflationary observables. Hence our practical shortcut turns to the usage of off-shell gauge-invariant effective action, $\Delta \Gamma \equiv0$.

Clearly (\ref{DeltaWeq}) and (\ref{key3}) show that the variation of the off-shell effective action, under a change of the gauge choice, originates from the variation of the fields $J^i\delta_{\scaleto{\rm  BRST}{3pt}} \varphi_i\propto J^i\delta_\alpha\varphi_i$ in the current terms. Therefore, as it is also clear by (\ref{key3}), that if the current term is symmetric, $i.e.$, $J^i\delta_\alpha\varphi_i=0$, $W$ and $\Gamma$ are invariant under gauge transformation. 

For example, in the case of $U(1)$ model, we can choose polar coordinates with the current-field coupling $J_\rho \rho$, which is clearly gauge invariant, gives 
\begin{eqnarray} 
\Delta W =- \int d^4x J_\rho(x)\left\langle\delta_\alpha\rho(x)\eta^\alpha\bar{\eta}_A\Delta\mathcal{F}^A\right\rangle= 0~,
\end{eqnarray} 
since $\rho$ is a gauge singlet and $\delta_\alpha\rho(x)=0$. This leads to an off-shell gauge-invariant effective action. Note that perturbativity is not assumed in this formal derivation, and the off-shell gauge invariance should be valid non-perturbatively, as a direct consequence of symmetry. This shows that an easy and practical way to achieve gauge independence is to use a polar-like current term to lift up the system away from the minimum.

\section{The unitary limit}\label{unitaryLim}
In addition to the straightforward way of adding a gauge-invariant current term, there is another practical way to approach off-shell gauge-invariance, namely, one can take the unitary gauge limit. For an general choice of current, the symmetry of the system might not be preserved, and the effective potential is manifestly dependent on the gauge-fixing parameter $\xi$. Inspired by \cite{Dolan:1974gu,Tye:1996au}, we can check what role the unitary gauge plays in the flow equations. That is, how the apparent gauge-dependence of the effective potential behaves for large values of the gauge parameter. By checking the behavior of Nielsen's flow equation at large $\xi$, we shall show that as $\xi$ increases, the gauge dependence introduced through an asymmetric background current is weakened,
and even insignificant for $\xi$ large enough.
Moreover, we show that in the unitary limit it is independent of the arbitrary external current, which can explain why the unitary gauge result exactly matches the gauge-invariant current result.

\begin{figure}[h!]
	\centering
	\includegraphics[width=12cm]{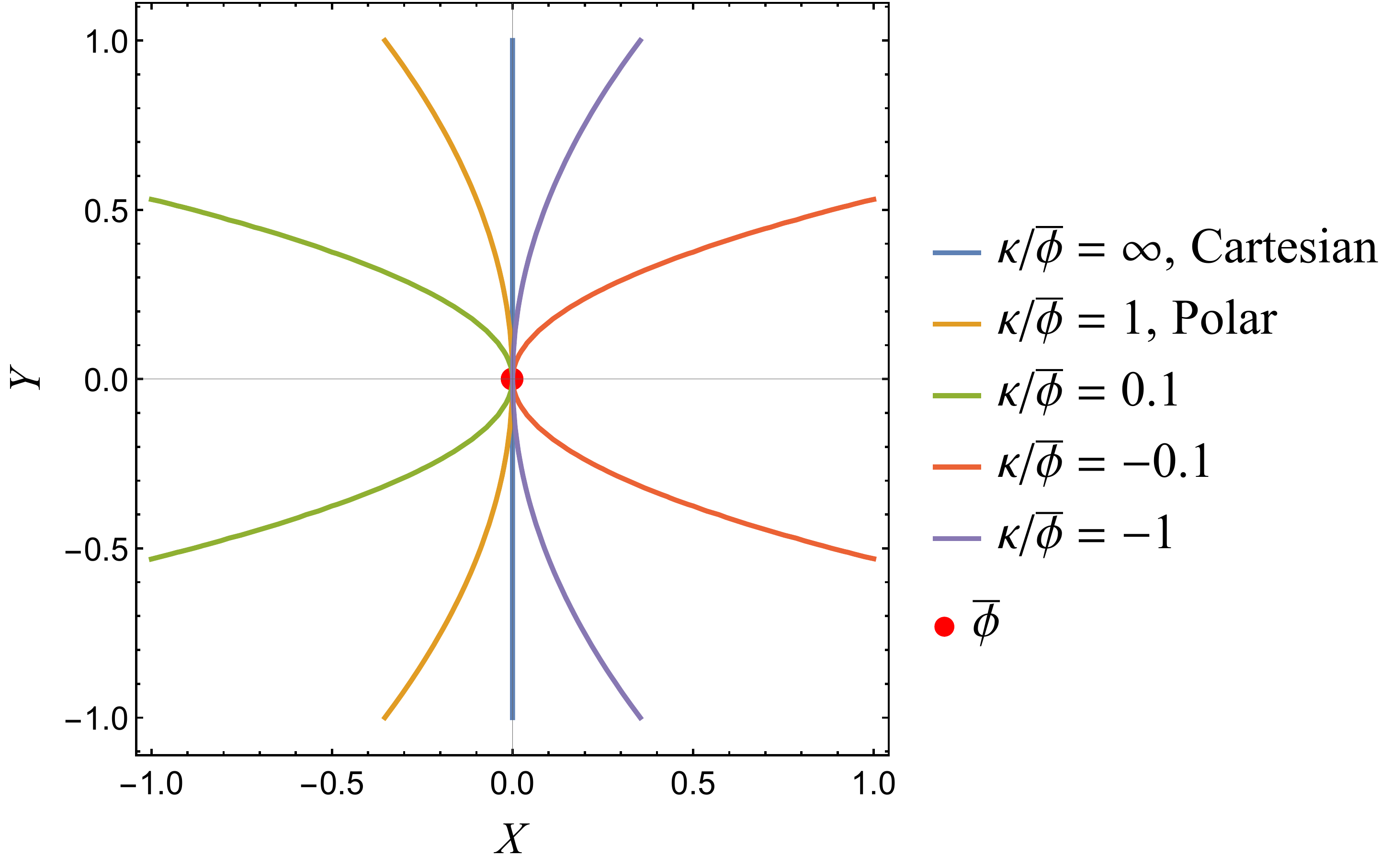}
	\caption{The $\kappa$ parameter stands for different choices of local coordinate systems on $\mathbb{R}^2$ spanned by $X,Y$. For a given choice of coordinate, the Goldstone mode fluctuations across the average field $\bar{\phi}$ along a constant Higgs line, whose curvature radius is given by $\kappa$. In this figure, we have shown different constant Higgs lines with different choices of $\kappa$.}\label{kappaIllustration}
\end{figure}

For simplicity, we use the $U(1)$ model quantized in the $R_\xi$ gauge as a demonstration,
\begin{equation}
	\mathcal{L}=-\frac{1}{4}F^2+|D\phi|^2-V(|\phi|)-\frac{1}{2\xi}(\partial\cdot A-\sqrt{2}e\xi\bar{\phi}Y)^2+\mathcal{L}_{\text{ghost}}~,\label{AHiggsLagrangian}
\end{equation}
By expanding $\phi$ around a background $\phi=\bar{\phi}+(X+iY)/\sqrt{2}$, we extract the Lagrangian up-to quadratic order as
\begin{eqnarray}
	\nonumber\mathcal{L}_2&=&-\frac{1}{2}X\left(\partial^2+6\lambda\bar{\phi}^2-m^2\right)X\\
	\nonumber&&-\frac{1}{2}Y\left(\partial^2+2\lambda\bar{\phi}^2-m^2+2\xi e^2\bar{\phi}^2\right)Y\\
	\nonumber&&+\frac{1}{2}A_\mu\left[(\partial^2+2e^2\bar{\phi}^2)\eta^{\mu\nu}-\left(1-\frac{1}{\xi}\right)\partial^\mu\partial^\nu\right]A_\nu\\
	\nonumber&&-\bar{c}(\partial^2+2\xi e^2\bar{\phi}^2)c\\
	&&-\sqrt{2}\bar{\phi}(2\lambda\bar{\phi}^2-m^2)X~.
\end{eqnarray}
The last term in the quadratic Lagrangian is linear and represents a tadpole that must be removed by the external current via the steepest descent method. We shall use a one-parameter family of external current $J_\kappa(\bar{\phi}+\frac{X}{\sqrt{2}}+\frac{Y^2}{4\kappa}+\mathcal{O}(XY^2))$, where $\kappa$ controls the background current adapted to different local coordinate systems on the scalar manifold. The physical meaning of $\kappa$ is the curvature radius of a constant Higgs line across $\bar{\phi}$, see Fig.~\ref{kappaIllustration}. The disappearance of an $X^2$ term is because we require the Higgs mass to be unchanged. Another way of viewing this is because we do not wish to further confine the fields around $\bar{\phi}$ in the $X$ direction by the external current.  The removal of the tadpole is accompanied by a change in the Goldstone mass. For a $U(1)$-symmetric current $\kappa=\bar{\phi}$, the current term becomes $J_\rho \rho$ in disguise, and the Goldstone mass is canceled as shown in the Sect.~\ref{U1Examp}, leading to a gauge-independent effective potential. However, if one uses an asymmetric current $\kappa\neq\bar{\phi}$ (in particular, $\kappa\rightarrow\infty$ is the conventional Cartesian current), the Goldstone remains massive, leading to an effective potential that does depend on $\xi$ and $\kappa$:
\begin{eqnarray}\label{RxiEffPotential}
	\nonumber V_{\text{eff}}(\bar{\phi},\xi)|_\kappa&=&V(\bar{\phi})-\frac{i}{2VT}\Bigg\{
	\Tr\ln\left[\partial^2+6\lambda\bar{\phi}^2-m^2\right]\\
	\nonumber&&~~~~~~~~~~~~~~~~~+\Tr\ln\left[\partial^2+(2\lambda\bar{\phi}^2-m^2)\left(1-\frac{\bar{\phi}}{\kappa}\right)+2\xi e^2\bar{\phi}^2\right]\\
	\nonumber&&~~~~~~~~~~~~~~~~~+\Tr\ln\left[-(\partial^2+2e^2\bar{\phi}^2)\eta^{\mu\nu}+\left(1-\frac{1}{\xi}\right)\partial^\mu\partial^\nu\right]\\
	&&~~~~~~~~~~~~~~~~~-2\Tr\ln\left[\partial^2+2\xi e^2\bar{\phi}^2\right]\Bigg\}+\text{(counterterms)}~.
\end{eqnarray}

We wish to use Nielsen's identity (\ref{NielsenId}) to show that the $\xi$-dependence of $V_{\text{eff}}$ is weakened in the large-$\xi$ limit, provided that the potential gradient $|\frac{\partial V_{\text{eff}}}{\partial\bar{\phi}}|$ is bounded as $\xi\rightarrow\infty$, which is indeed the case for the one-loop potential (\ref{RxiEffPotential}).

In this $U(1)$ model, the $C(\bar{\phi},\xi)$ function in Nielsen's identity (\ref{NielsenId}) is given by
\begin{equation}
	C(\bar{\phi},\xi)=\frac{1}{\delta\ln\xi}\left\langle\frac{\delta X}{\sqrt{2}}+\frac{2Y\delta Y}{4\kappa}+(\text{higher order in fields})\right\rangle\Big|_{\bar{\phi}}~.
\end{equation}
Here $\delta X$ and $\delta Y$ are a particular field redefinition that takes the form of a gauge transformation which preserves (\ref{KeepFeq}) after sending the gauge parameter from $\xi\rightarrow\xi+\delta\xi$. From (\ref{KeepFeq}), it is not difficult to find the required field redefinition for $R_\xi$ gauge choices is
\begin{eqnarray}
	\delta X&=&-\frac{e}{2}Y\frac{1}{\partial^2+2\xi e^2\bar{\phi}^2}\left(\partial\cdot A\delta \ln\xi+\sqrt{2}e\bar{\phi}Y\delta\xi\right)\\
	\delta Y&=&\frac{e}{2}(\sqrt{2}\bar{\phi}+X)\frac{1}{\partial^2+2\xi e^2\bar{\phi}^2}\left(\partial\cdot A\delta \ln\xi+\sqrt{2}e\bar{\phi}Y\delta\xi\right)~.
\end{eqnarray}
These are essentially the BRST transformation $\delta_{\scaleto{\rm  BRST}{3pt}}\varphi_i$ in (\ref{key3}) non-localized after integrating out the ghost field. 

Therefore by the Nielsen identity, the dependence of the effective potential on the gauge parameter is given by
\begin{equation}
	\frac{\partial V_{\text{eff}}}{\partial\ln\xi}=-\frac{e}{2\sqrt{2}}\frac{\partial V_{\text{eff}}}{\partial\bar{\phi}}\left(1-\frac{\bar{\phi}}{\kappa}\right)\left\langle Y\frac{1}{\partial^2+2\xi e^2\bar{\phi}^2}\left(\partial\cdot A+\sqrt{2}\xi e\bar{\phi}Y\right)\right\rangle\Bigg|_{\bar{\phi}}~,\label{largexiVdxi}
\end{equation}
where the higher order terms in fields are omitted since they are also higher order in $\xi^{-1}$ than the leading order. Notice that for a gauge-invariant current choice, $\kappa=\bar{\phi}$ and the effective potential is directly independent\footnote{This is for the leading order. Actually if one consider the full expansion of $\rho$ into $X,Y$ fields, the $C$ function vanish order by order in the $X,Y$ series, since $\rho$ itself is to all orders a gauge singlet.} of $\xi$, as shown in (\ref{key3}). In the large $\xi$ limit,
\begin{equation}
\left\langle Y\frac{1}{\partial^2+2\xi e^2\bar{\phi}^2}\left(\partial\cdot A+\sqrt{2}\xi e\bar{\phi}Y\right)\right\rangle\Bigg|_{\bar{\phi}}\xrightarrow{\xi\rightarrow\infty,\text{ fixing cutoff}}\frac{1}{\sqrt{2}e\bar{\phi}}\langle Y^2\rangle=\mathcal{O}(\xi^{-1})~.\label{largexiC}
\end{equation}
Therefore, if $|\frac{\partial V_{\text{eff}}}{\partial\bar{\phi}}|<\mathcal{O}(\xi^{0})$, the gauge dependence vanishes in the unitary limit, $|\frac{\partial V_{\text{eff}}}{\partial\ln \xi}|\lesssim \mathcal{O}(\xi^{-1})\rightarrow 0$. Indeed, the one-loop result behaves as
\begin{eqnarray}
	\nonumber\frac{\partial V_{\text{eff}}}{\partial\bar{\phi}}&=&V'(\bar{\phi})-\frac{i}{2VT}\Bigg\{\Tr\frac{12\lambda\bar{\phi}}{\partial^2+6\lambda\bar{\phi}^2-m^2}\\
	\nonumber&&~~~~~~~~~~~~~~~~~~+\Tr\frac{(4\lambda+4\xi e^2)\bar{\phi}-(6\lambda\bar{\phi}^2-m^2)/\kappa}{\partial^2+\left(2\lambda\bar{\phi}^2-m^2\right)\left(1-\bar{\phi}/\kappa\right)+2\xi e^2\bar{\phi}^2}\\
	\nonumber&&~~~~~~~~~~~~~~~~~~+\Tr\left[\left(-(\partial^2+2e^2\bar{\phi}^2)\eta^{\mu\nu}+\left(1-\frac{1}{\xi}\right)\partial^\mu\partial^\nu\right)^{-1}\left(4e^2\bar{\phi}\eta^{\nu\rho}\right)\right]\\
	\nonumber&&~~~~~~~~~~~~~~~~~~-2\Tr\frac{4\xi e^2\bar{\phi}}{\partial^2+2\xi e^2\bar{\phi}^2}\Bigg\}\\
	&=&\text{gauge independent terms}+\frac{i}{2VT}\Tr\frac{4\xi e^2\bar{\phi}}{\partial^2+2\xi e^2\bar{\phi}^2} + \mathcal{O}(\xi^{-1})~.\label{largexiVdphi}
\end{eqnarray}
Holding the cut off fixed, we send $\xi$ to infinity\footnote{Note that this is necessary since the purpose of sending $\xi$ to infinity is to kill the Goldstone degree of freedom, hence the unitary limit. If we renormalize and send the cutoff to infinity first, then however large $\xi$ is, at a sufficiently high energy scale in the loop integral, the Goldstone degree of freedom will enter and bring the gauge dependence along into the result.}, the second term becomes independent of $\xi$. Therefore, $V'_{\text{eff}}$ is bounded by a $\xi$-independent constant $\mathcal{O}(\xi^0)$. 

As a result, combining (\ref{largexiVdxi}), (\ref{largexiC}) and (\ref{largexiVdphi}), the dependence of $V_{\text{eff}}$ on the gauge parameter is given by
\begin{equation}
	\frac{\partial V_{\text{eff}}}{\partial\ln\xi}\xrightarrow{\xi\rightarrow\infty,\text{ fixing }\kappa\text{ \& cutoff}}\mathcal{O}(\xi^{-1})~.
\end{equation}
Thus, taking the limit $\xi\rightarrow\infty$, we see that the effective potential is independent of any $\xi$-rescaling in this unitary limit.

To show the independence of the effective potential on the current choice $\kappa$, we take the derivative of (\ref{RxiEffPotential}) with respect to $\ln\kappa$:
\begin{equation}
	\frac{\partial V_{\text{eff}}}{\partial\ln \kappa}=-\frac{i}{2VT}\Tr\frac{\left(2\lambda\bar{\phi}^2-m^2\right)\bar{\phi}/\kappa}{\partial^2+\left(2\lambda\bar{\phi}^2-m^2\right)\left(1-\bar{\phi}/\kappa\right)+2\xi e^2\bar{\phi}^2}\xrightarrow{\xi\rightarrow\infty,\text{ fixing }\kappa\text{ \& cutoff}}\mathcal{O}(\xi^{-1})~.
\end{equation}
Thus the dependence on the current choice also vanishes in the unitary limit. In this sense, we consider unitary gauge result as the fixed point under the gauge flow of changing $\xi$ defined by the Nielsen identity (\ref{NielsenId}). It is invariant under both the change of $\xi$ (gauge choice) and $\kappa$ (current choice), and is what the gauge-dependent effective potential converges to.

\section{Conclusion}\label{conclusion}
We have given an explicit demonstration of the apparent gauge dependence issue in Higgs-like inflation scenario, namely its \textit{non-negligible} influence on the $n_s$-$r$ prediction of inflation. We qualitatively analyzed the possible underlying causes and found it might be due to either a truncation of the non-perturbative effective action to second order in gradients, or curved spacetime effects in the loop diagram evaluation of the effective action. To resolve this problem, one can either seek more rigorous but complicated ways in the conventional formalism, or find out -as we did- how exactly the gauge dependence infiltrates the effective potential and construct an off-shell gauge independent one.

We found that if one chooses as fundamental fields some field space coordinates that give a current term which violates gauge symmetry, the resulting effective potential is spoiled. Therefore, we have used as fundamental fields those whose corresponding current terms are gauge-invariant. This is equivalent to choosing polar coordinates on the scalar manifold. In this way we have obtained an off-shell gauge-independent effective potential. After working out various examples and giving our $n_s$-$r$ prediction for the Abelian model, we used Nielsen's identity to show why it works even non-perturbatively. At last we discussed the relation between the polar coordinates result and the unitary gauge result. We found that in accordance with Nielsen's identity, both the gauge dependence and the current choice dependence is weakened in the unitary limit and the resulting effective potential all converge to that of the polar coordinates.

Finally we would like to mention that our method, being a practical shortcut, faces several challenges.
\begin{enumerate}
	\item[$\bullet$] Our method is different from the mainstream approach.
	Therefore, even though it gives a gauge independent result, a key challenge is to prove that this is \textit{the} right way to get to correct physical predictions.
	\item[$\bullet$] Second is the integration of the $\rho$ field, which we treat as a Gaussian, whereas the integral bounds make it an error function. Our key assumption here is that the potential at each point receives significant corrections from neighboring points only. Moreover, as we are here interested in Higgs inflation, we expect large values of $\rho$ rather than inflating near the origin, where polar coordinates are ill-defined. Large field values come with large second derivatives $V''(\rho)$, hence leading to a restriction localizing $\rho$. This kind of field-space locality needs further elaboration to be proven a good approximation. It is good to note that such a difficulty does not arise in the unitary gauge.
	\item[$\bullet$] The third shortcoming of our method is that for the angular-field integrals, the truncation to second order might be questionable, especially for non-Abelian theories. This is due to the existence of terms in the Lagrangian (e.g. schematically, terms such as $ \rho^2( \partial \theta)^2$ in the scalar QED case, and $ \frac{\bar{\phi}^2\sin^2 \theta}{\theta^2} ( \partial \theta)^2 $ in the $SU(2)$ case) which are not suppressed at higher loop orders, and the geometry of the coset manifold is not taken into account here by truncating to second order in angular variables. Therefore, although the symmetric polar current term does ensure the gauge independence of the effective potential up to all loop orders, in practice, one still cannot directly enjoy this gauge independence to the fullest because of these uncontrolled non-renormalizable terms.
\end{enumerate}
As a result, the true accuracy of our method should in the end be determined by comparison to more rigorous first-principle computations or future experiments. Despite the weaknesses, which we have left for future work, our method is highly practical for cosmological considerations.

\section*{Acknowledgment} 
	We would like to thank Henry Tye for guidance and initial collaboration. We also thank Andrew Cohen for helpful discussions, comments, and constructive criticism. We are indebted to Kun-feng Lyu, Yi Wang and Haitham Zaraket for their valuable suggestions on our manuscript.
\appendix

\section{Effective potential in Abelian Higgs inflation}\label{AHiggsInfAppendix}
\subsection{Cartesian field coordinates}
We first work in the conventional Cartesian field coordinates. In the Jordan frame, we neglect the spacetime curvature and approximate $g_{\mu\nu}\approx \eta_{\mu \nu}$. Then the Abelian Higgs model (\ref{AHiggsLagrangian}) in the $R_\xi$ gauge yields a $\xi$-dependent effective potential regularized using the $\overline{\text{MS}}$-scheme,
\begin{eqnarray}
	\nonumber V_{\text{eff}}(\bar{\phi},\xi)&=&-m^2\bar{\phi}^2+\lambda\bar{\phi}^4\\
	&&+\frac{1}{4(4\pi)^2}\left\{\sum_{b=X,Y,A_i}m_b^4(\bar{\phi},\xi)\left[\ln \frac{m_b^2(\bar{\phi},\xi)}{M^2}-\frac{3}{2}\right]-m_c^4(\bar{\phi},\xi)\left[\ln\frac{m_c^2(\bar{\phi},\xi)}{M^2}-\frac{3}{2}\right]\right\}~,~~~~~~\label{CartesianVeff}
\end{eqnarray}
where
\begin{eqnarray}
	\nonumber m_X^2&=&6\lambda\bar{\phi}^2-m^2\\
	\nonumber m_Y^2&=&2\lambda\bar{\phi}^2-m^2+2\xi e^2\bar{\phi}^2\\
	\nonumber m_A^2&=&2e^2\bar{\phi}^2\\
	m_c^2&=&2\xi e^2\bar{\phi}^2~,
\end{eqnarray}
and $M$ is an arbitrary scale to be determined by renormalization. Note that in this appendix we drop subscript $J$ in the classical field $\bar{\phi}_J$ for convenience and $\bar{\phi}$ is understood to be held by a background current. For simplicity, let us consider the massless case $m=0$. And we impose the renormalization condition for $V_{\text{eff}}$ as
\begin{equation}
	\frac{\partial^4V_{\text{eff}}}{\partial\bar{\phi}^4}\Big|_{\bar{\phi}=\mu}=4!\lambda~.\label{lambdaRenormCond}
\end{equation}
Solving $M$ in terms of $\mu$, we obtain
\begin{equation}
	V_{\text{eff}}(\bar{\phi},\xi)=\lambda\bar{\phi}^4+\frac{10\lambda^2+3e^4+2\xi\lambda e^2}{(4\pi)^2}\bar{\phi}^4\left[\ln\frac{\bar{\phi}^2}{\mu^2}-\frac{25}{6}\right]~,\label{VeffCartesianLRxiRenormed}
\end{equation}
in agreement with \cite{Coleman:1973jx} in the Landau gauge $\xi\to 0$.

\begin{figure}[h!]
	\centering
	\includegraphics[width=17cm]{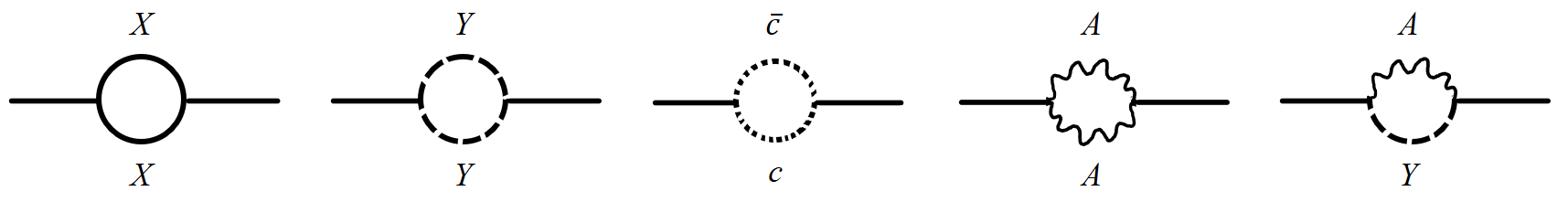}
	\caption{The loop diagrams contributing to the finite part of $Z(\bar{\phi},\xi)$ in Cartesian coordinates, which is dominated by the gauge-boson-Goldstone loop in the parameter regime we are interested in.}\label{Zedloops}
\end{figure}
The field strength factor $Z(\bar{\phi},\xi)$ receives contributions from five diagrams shown in Fig.~\ref{Zedloops}, of which the dominating diagram is the logarithmically divergent gauge-boson-Goldstone loop. Taking their momentum derivatives at zero external momenta, we obtain the $\overline{\text{MS}}$-scheme field strength factor 
\begin{eqnarray}
\nonumber Z-1&=&\frac{1}{(4\pi)^2}\Bigg\{\frac{6 \lambda ^2 \bar\phi
		^2}{m_X^2}+\frac{2 \lambda ^2 \bar\phi ^2}{3m_Y^2}-\frac{\xi e^4\bar{\phi}^2}{3m_A^2}+\frac{e^4 \bar\phi ^2 \left(11 \xi ^2+4 \xi -18 \xi  \ln\xi-15\right)}{3(\xi -1) m_A^2}\\
\nonumber&&~~~~~~~~~~+\frac{e^2 \xi ^2 m_A^2 \left(\xi ^2 m_A^4-\xi  m_A^2 m_Y^2-m_Y^4\right)\ln\xi}{\left(m_Y^2-\xi 
			m_A^2\right){}^3}\\
\nonumber&&~~~~~~~~~~+\frac{e^2}{\left(m_A^2-m_Y^2\right)
			\left(m_Y^2-\xi m_A^2\right)^3}\\
\nonumber&&~~~~~~~~~~~~~\times\Big[m_A^2 \ln\frac{m_A^2}{M^2}\Big((\xi -3)\xi^3 m_A^6-\xi ^2 \left(\xi ^2+\xi -9\right) m_A^4 m_Y^2\\
\nonumber&&~~~~~~~~~~~~~~~~~~~~~~~~~~~~~~~~+\xi((\xi-1)\xi-9) m_A^2 m_Y^4+\left(\xi
	^2+3\right) m_Y^6\Big)\\
\nonumber&&~~~~~~~~~~~~~~~+m_Y^2 \ln\frac{m_Y^2}{M^2}\Big(\xi ^3 m_A^6+\xi ^2 (2 \xi -5) m_A^4 m_Y^2\\
\nonumber&&~~~~~~~~~~~~~~~~~~~~~~~~~~~~~~~~-4 (\xi -2) \xi  m_A^2 m_Y^4+(\xi -3) m_Y^6\Big)\Big]\Bigg\}~.\\
\end{eqnarray}
In the massless case, we impose the renormalization condition
\begin{equation}
	Z\Big|_{\bar{\phi}=\mu}-1=0\label{ZRenormCond}
\end{equation}
and solve $M$ in terms of $\mu$. The result simplifies considerably,
\begin{equation}
	Z(\bar{\phi},\xi)=1+\frac{e^2}{(4\pi)^2}(3-\xi)\ln\frac{\bar{\phi}^2}{\mu^2}~,\label{ZedLRxiRenormed}
\end{equation}
in agreement with \cite{Nielsen:2014spa}.

The running of the non-minimal coupling term $\alpha|\phi|^2R$ receives contribution from both the flat-spacetime renormalization of the $|\phi|^2$ operator and a new renormalization constant due to spacetime curvature \cite{Buchbinder:1992rb}. However, at one-loop level, these two contributions are related by a simple formula \cite{Buchbinder:1985ba}. Hence one can rewrite the one-loop running of $\alpha$ parameter entirely in terms of the anomalous dimension of $|\phi|^2$ \cite{Buchbinder:1985ew},
\begin{equation}
	\frac{d\alpha_{\text{eff}}}{d \ln(\bar\phi/\mu)}=\left(\alpha_{\text{eff}}+\frac{1}{6}\right)\gamma_{|\phi|^2},~~~\alpha_{\text{eff}}|_{\bar{\phi}=\mu}=\alpha~,
\end{equation}
where we have imposed the renormalization condition. In the Cartesian-coordinate case, after reading off $\gamma_{|\phi|^2}$ from the effective potential (\ref{CartesianVeff}), we can easily solve the effective non-minimal coupling up to one-loop order as
\begin{equation}
	\alpha_{\text{eff}}(\bar{\phi},\xi)=\alpha+\left(\alpha+\frac{1}{6}\right)\frac{4\lambda+\xi e^2}{(4\pi)^2}\ln \frac{\bar{\phi}^2}{\mu^2}~.
\end{equation}
Again, the running non-minimal coupling is also gauge-dependent. This running non-minimal coupling will affect the final effective potential when mapping back to the Einstein frame.

\subsection{Polar field coordinates}
In the polar-coordinate case, $\phi=(\bar{\phi}+\rho)e^{i\theta}$, we also work in the $R_\xi$ gauge family and add a gauge-fixing term $-\frac{1}{2\xi}(\partial\cdot A-e\bar \phi ^2 \xi\theta)^2$. the effective potential looks the same as (\ref{CartesianVeff}) but without the Goldstone and ghost terms,
\begin{eqnarray}
\nonumber V_{\text{eff,~polar}}(\bar{\phi},\xi)&=&-m^2\bar{\phi}^2+\lambda\bar{\phi}^4\\
&&+\frac{1}{4(4\pi)^2}\sum_{b=X,A_i}m_b^4(\bar{\phi},\xi)\left[\ln \frac{m_b^2(\bar{\phi},\xi)}{M^2}-\frac{3}{2}\right]~,~~~~~~\label{PolarVeff}
\end{eqnarray}
Using the same renormalization condition (\ref{lambdaRenormCond}), the massless limit gives
\begin{equation}
V_{\text{eff,~polar}}(\bar{\phi},\xi)=\lambda\bar{\phi}^4+\frac{9\lambda^2+3e^4}{(4\pi)^2}\bar{\phi}^4\left[\ln\frac{\bar{\phi}^2}{\mu^2}-\frac{25}{6}\right]~.
\end{equation}
\begin{figure}[h!]
	\centering
	\includegraphics[width=14cm]{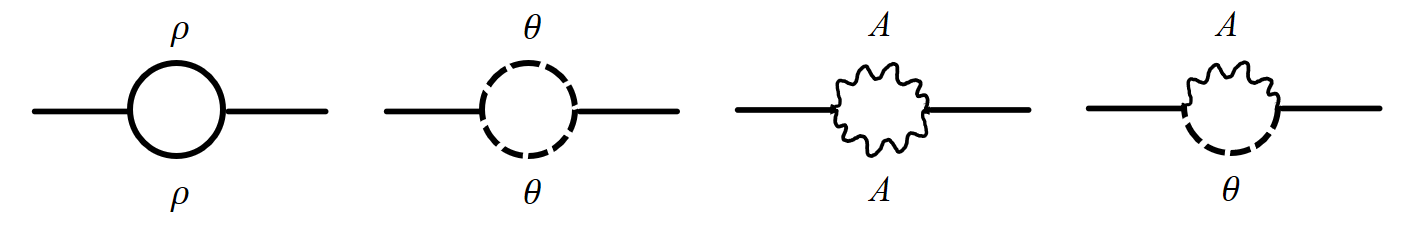}
	\caption{The loop diagrams contributing to the finite part of $Z_\text{polar}(\bar{\phi},\xi)$ in polar coordinates. There is no ghost loop because the polar-coordinate ghost is free and does not couple to the $\rho$ field.}\label{ZedloopsPolar}
\end{figure}
The field-strength factor receives contribution from four diagrams shown in Fig.~\ref{ZedloopsPolar}. After applying the renormalization condition (\ref{ZRenormCond}), we obtain
\begin{equation}
Z_\text{polar}(\bar{\phi},\xi)=1+\frac{3e^2}{(4\pi)^2}\ln\frac{\bar{\phi}^2}{\mu^2}~
\end{equation}
in the massless case. The agreement of $Z_\text{polar}(\bar{\phi},\xi)$ with that computed in the covariant gauge family in \cite{Tye:1996au} shows again the gauge-family-independence of polar-coordinate quantities.

The running non-minimal coupling evaluated in the polar field coordinates is similarly obtained from the mass renormalization at one-loop order. The result is, again, gauge-independent:
\begin{equation}
	\alpha_{\text{eff, polar}}(\bar{\phi},\xi)=\alpha+\left(\alpha+\frac{1}{6}\right)\frac{3\lambda}{(4\pi)^2}\ln \frac{\bar{\phi}^2}{\mu^2}~.
\end{equation}

\subsection{Going into the Einstein frame}

Now, the combined field redefinition (\ref{ZJordanToEinstein}) can be performed separately in two steps. Namely we first compute the canonical field in Jordan frame, taking into account of the radiative corrections, and then we go to the Einstein frame, taking into account the running non-minimal coupling. 

\begin{figure}[h!]
	\centering
	\includegraphics[width=12cm]{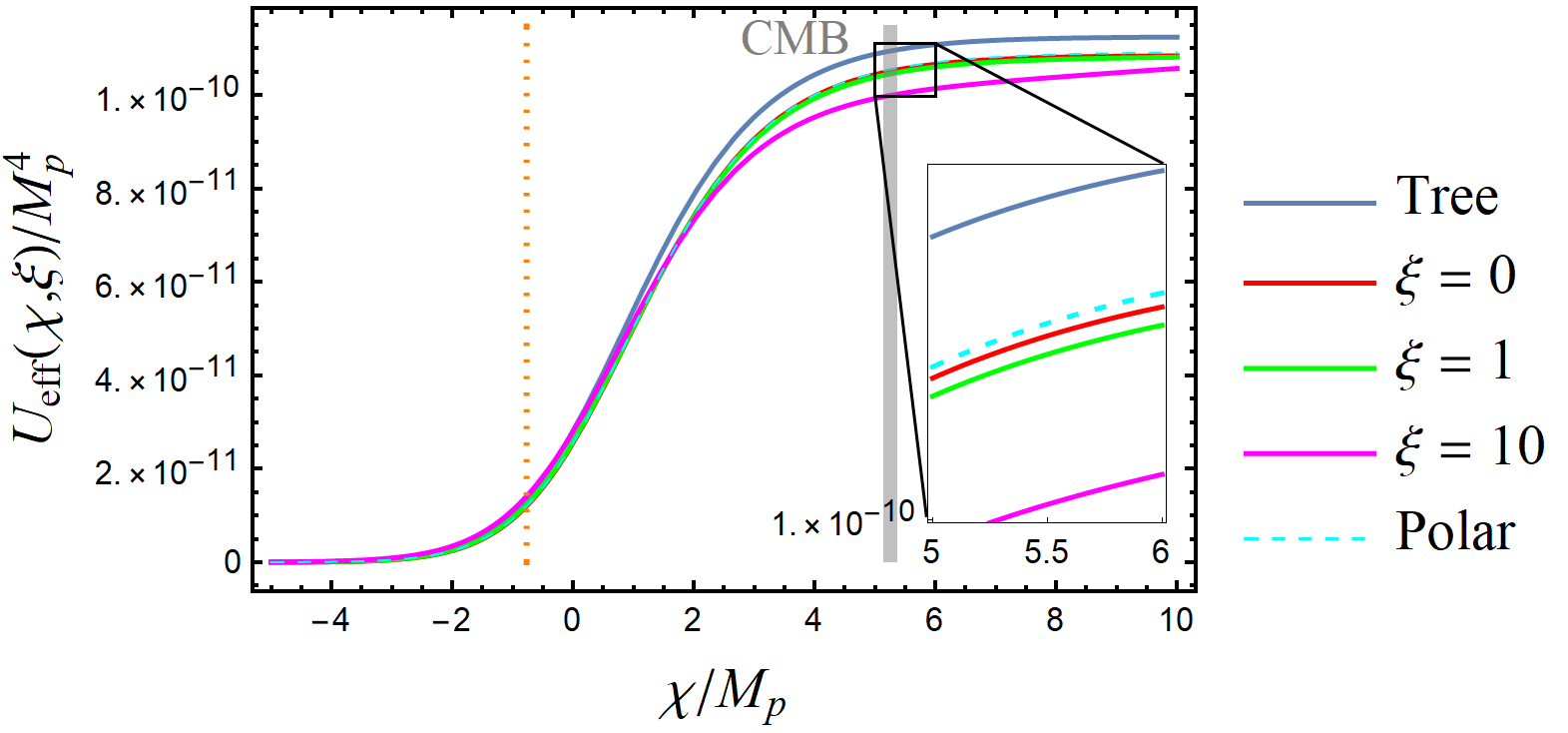}
	\caption{The shape of the final potential $U_{\text{eff}}(\chi,\xi)$ for different choices of $\xi$ in Cartesian coordinates (solid lines) and the polar one (dashed line). The parameters are chosen to be the same as in Fig.~\ref{ns-r-xiDep}, namely $\lambda=0.13, e=0.3, m=0, \alpha=17000,\mu/M_p=0.049$. The CMB scale with $50\lesssim N_{\text{CMB}}\lesssim60$ corresponds to the field range $5.1\lesssim \chi/M_p\lesssim 5.4$, and is indicated by the gray band. The dotted orange line represents the end of inflation defined by the exit condition $\epsilon=\mathcal{O}(1)$.}\label{UeffShape}
\end{figure}

The canonical field in Jordan frame is given by $\bar{\phi}_c=\sqrt{2}\int Z(\bar{\phi},\xi)^{1/2} d\bar{\phi}$. For all reasonable parameter choices, $|\delta Z|=|Z-1|\lesssim \frac{e^2}{(4\pi)^2}\times\mathcal{O}(1)\ll 1$, thus we can approximate
\begin{equation}
	\bar{\phi}=\frac{1}{\sqrt{2}}\left[\bar{\phi}_c-\int_{\sqrt{2}\mu}^{\bar\phi_c} \frac{1}{2}\delta Z(y/\sqrt{2},\xi) dy\right]~.\label{ZedCanon}
\end{equation}
The effective potential expressed in the canonical field $\bar{\phi}_c$ becomes then $V_{\text{eff}}(\bar{\phi}(\bar{\phi}_c,\xi),\xi)$ through (\ref{ZedCanon}). After canonicalizing the Higgs field in the Jordan frame, the second step is going to the Einstein frame by a field redefinition
\begin{equation}
\chi=\int \sqrt{\frac{\Omega^2+6\alpha_{\text{eff}}^2\bar{\phi}_c^2/M_p^2}{\Omega^4}}d\bar{\phi}_c\approx \sqrt{6}M_p\ln\frac{\sqrt{\alpha_{\text{eff}}}\bar{\phi}_c}{M_p},~~\text{where }~~~\Omega^2=1+\frac{\alpha_{\text{eff}} \bar{\phi}_c^2}{M_p^2}~.\label{JordanToEinstein}
\end{equation}
Here we have approximated the integral by considering our assumed parameter regime $\Omega\gg 1$ and $1\ll\sqrt{\alpha}\ll 10^{17}$ (see \cite{Bezrukov:2007ep} for more discussions). The shape of the final potential $U_{\text{eff}}(\chi,\xi)=\frac{V_{\text{eff}}(\bar{\phi}(\bar{\phi}_c(\chi,\xi),\xi),\xi)}{\Omega^4}$ is shown in Fig.~\ref{UeffShape}. 

Notice that in principle, a Renormalization Group (RG) improvement of $V_{\text{eff}}$ is still needed to resum potentially large logarithms. In this work, for simplicity, we did not perform such an analysis. Instead, we try to avoid large logarithms by renormalizing the system around its typical scale during inflation, namely the CMB scale lying at $\frac{\chi}{M_p}\sim 5.4\Leftrightarrow \frac{\bar\phi}{M_p}\sim 0.049$. Thus if we set the renormalization scale to be $\frac{\mu}{M_p}=0.049$, the field value during inflation is kept around $\bar\phi\sim \mu$. Since RG improvement starts only at two-loop order, we expect the error to be negligible as long as we stay in the perturbative regime and keep the logarithms small. In fact, the error can be easily estimated. During the last 60 e-folds of inflation, the field traverse $\Delta\chi\approx 6.1$ corresponds to a logarithm lying withing $-5<\ln\frac{\bar{\phi}^2}{\mu^2}<0$. Therefore the error is suppressed by an \textit{extra} factor of $\frac{\mathcal{O}(10\lambda,3e^2)}{(4\pi)^2}\times \left|\ln\frac{\bar{\phi}^2}{\mu^2}\right|< 5\times 10^{-2}$. A full RG investigation is left for future works. Notice, however, in similar models, gauge dependence still lurks around after RG improvement in Cartesian coordinates \cite{Cook:2014dga}.

At last, we point out that our computation is performed in the linear $R_\xi$ gauge with a gauge-fixing function $\mathcal{F}_L=\partial\cdot A-\sqrt{2}e\xi \bar{\phi} Y$. An alternative choice is the non-linear $R_\xi$ gauge with $\mathcal{F}_{NL}=\partial\cdot A-\sqrt{2}e\xi \left(\bar{\phi}+\frac{X}{\sqrt{2}}\right) Y$. In this gauge family, the Cartesian-coordinate effective potential remains the same as (\ref{VeffCartesianLRxiRenormed}). However, the Cartesian-coordinate anomalous dimension becomes different since the gauge-boson-Goldstone loop is modified. The outcome is a change of sign before the $\xi$ term in (\ref{ZedLRxiRenormed}). This sign change turns out to accidentally cancel the main gauge dependence in $V_{\text{eff}}$ after field canonicalization. This might seem to lead to the conclusion that canonicalization eliminates gauge dependence. It is, however, not true since the mere disagreement between linear/non-linear $R_\xi$ gauge is \textit{itself} another aspect of gauge dependence. Canonicalization does not solve the problem.

\bibliographystyle{utphys}
\bibliography{reference}

\end{document}